\documentclass[aps,prd,preprint,tightenlines,showpacs,nofootinbib,eqsecnum,amsfonts,amssymb,amsmath]{revtex4}
\usepackage{graphicx,epsfig,rotate,psfig}   % Include figure files
\usepackage{dcolumn}      % Align table columns on decimal point

\newcommand{\etal}{{\it et al.}}
\newcommand{\dd}{{\rm d}}
\newcommand{\cav}{\mathrm{cav}}

\newcommand{\ppn}{\mathrm{PPN}}

\newcommand{\mat}{\mathrm{mat}}
\newcommand{\rad}{\mathrm{rad}}
\newcommand{\scalefac}{R}
\newcommand{\baryon}{{\rm b}}
\newcommand{\xout}{{\rm out}}
\newcommand{\xin}{{\rm in}}
\newcommand{\xeq}{{\rm eq}}
\def\iso#1#2{\mbox{${}^{#2}{\rm #1}$}}

\def\he#1{\iso{He}{#1}}
\def\li#1{\iso{Li}{#1}}
\def\be#1{\iso{Be}{#1}}

\begin{document}
\leftline{UMN--TH--2721/08, FTPI--MINN--08/39}

\title{Non-universal scalar-tensor theories and big bang nucleosynthesis}

\author{Alain Coc}
 \email{coc@csnsm.in2p3.fr}
 \affiliation{Centre de Spectrom\'etrie Nucl\'eaire et de
Spectrom\'etrie de Masse, IN2P3/CNRS/UPS, B\^at. 104, 91405 0rsay
Campus (France)}

\author{Keith A. Olive}
 \email{olive@physics.unm.edu}
 \affiliation{William I. Fine Theoretical Physics Institute,
              University of Minnesota, Minneapolis, MN 55455 (USA)}

\author{Jean-Philippe Uzan}
 \email{uzan@iap.fr}
 \affiliation{Institut d'Astrophysique de Paris,
              UMR-7095 du CNRS, Universit\'e Pierre et Marie
              Curie,
              98 bis bd Arago, 75014 Paris (France)}

\author{Elisabeth Vangioni}
 \email{vangioni@iap.fr}
 \affiliation{Institut d'Astrophysique de Paris,
              UMR-7095 du CNRS, Universit\'e Pierre et Marie
              Curie,
              98 bis bd Arago, 75014 Paris (France)}

\begin{abstract}
We investigate the constraints that can be set from
big-bang nucleosynthesis on two classes of models: extended
quintessence and scalar-tensor theories of gravity in which the
equivalence principle between standard matter and dark matter is
violated. In the latter case, and for a massless dilaton with
quadratic couplings, the phase space of theories is investigated.
We delineate those theories where attraction toward general relativity occurs.
It is shown that big-bang nucleosynthesis sets more stringent constraints than
those obtained from Solar system tests.
\end{abstract}
 \date{6 aout 2008}
 \maketitle
%%%%%%%%%%%%%%%%%%%%%%%%%%%%%%%%%%%%%%%%%%%%%%%%%%%%%%%%%%%%%%%%%%%%%%%%%%%%%%
\section{Introduction}

While the late-time acceleration of the expansion of our universe
seems to be a robust conclusion reached by several independent
cosmological observations~\cite{SCP,obs1,obs2,dunkley,kilbinger},
its origin remains an open question~\cite{jpu1,jpugrg}.
Various explanations for late-time acceleration have been proposed
ranging from modifications of the laws of physics to questioning
the Copernican principle~\cite{goodman,pc1,pc2}. If this principle
holds we are lead to the conclusion that our cosmological model
requires the introduction of new physical degrees of freedom,
often referred to as the ``dark sector''.  In this case, it is necessary to
specify the interactions of these new fields, both their self interactions
as well as their interactions with matter.
There are several candidates for this new matter,  the simplest being quintessence
models~\cite{quint}, or a modification of general relativity (see
Ref.~\cite{jpugrg} for a discussion on this distinction).
Let us emphasize that the simplest model of all requires only the
introduction of a single constant, the cosmological constant, and
is at the moment in agreement with all available
data~\cite{kilbinger,schimdetal}.

Among all the theories of gravitation beyond general relativity,
scalar-tensor theories~\cite{stgen,def} are a well motivated
extension and may also be related to quintessence since they
include one or more scalar fields universally coupled to matter.
These theories are described by two functions, a potential
describing the scalar self interactions, and a coupling function
describing interaction of the scalar field to matter. Indeed,
these theories depart from general relativity and are subject to
many experimental constraints, particularly from the Solar
system~\cite{willlrr}. These deviations can however be larger on
cosmological scales because, as long as the coupling function has
a minimum, such theories can be dynamically attracted~\cite{dn}
toward general relativity during the evolution of the universe.
More generally, when considering theories with a quintessence-like
potential, it was shown~\cite{jpu99,bp} that the theory can be
attracted toward general relativity while the scalar field is
driven toward the quintessence tracking solution, hence leading to
models of extended quintessence that were advocated to be very
generic for quintessence
models~\cite{runawaydil,quintnonuniversal}. In that context, the
deviations from general relativity can be constrained from big
bang nucleosynthesis (BBN)~\cite{dp,couv}, the cosmic microwave
background (CMB)~\cite{ru}, weak lensing~\cite{sur04} and the
production of gravitational waves~\cite{ru2}. Interestingly, they
can also lead~\cite{msu06} to an equation of state with $p/\rho < -1$
for the dark energy.

It is indeed also possible that the new scalar field(s) do not
couple universally to matter, as considered in
Refs.~\cite{runawaydil,quintnonuniversal}. In this case, one
expects a violation of the equivalence principle and a variation of
the fundamental constants which can used to constrain the equation
of state of the dark energy~\cite{quintalpha}. However, such couplings are
in general severely constrained~\cite{jpuct,ctes2}, with the exception of
some particular cases such as the coupling to neutrinos~\cite{nu}
or dark matter.

When the scalar fields couple
differently to the dark sector, there is almost no constraint from
the equivalence principle on the interactions between dark matter (DM) and standard
(or visible) matter (see however Ref.~\cite{tidal} for a possible
test of the universality of free fall between DM and normal
matter). Such a possibility was initially investigated in
Ref.~\cite{dgg} in the particular case of Brans-Dicke theories and
was recently revisited in Refs.~\cite{alimi}. These models can, in
principle explain, the coincidence problem as
well~\cite{qinteract1,qinteract2,qinteract3} and have been argued
to naturally appear when the quintessence model is embedded in a
supersymmetric construction~\cite{bm}.

In this article, we focus on big bang nucleosynthesis and build on
our previous analysis~\cite{couv,cnouv} that focused on the
constraints that can be imposed on scalar-tensor theories of
gravity with a massless dilaton. We extend the results of
Ref.~\cite{couv} in two directions: (1) by considering extended
quintessence models, i.e. a general scalar-tensor theory with a
potential, and (2) by extending our formalism to models with a
non-universal coupling between the dark matter sector and the
visible sector. The theories considered in this article are first
detailed in \S~\ref{sec2} and both cases are studied respectively
in \S~\ref{sec3} and \S~\ref{sec4}. Most technicalities, such as
the BBN data and Solar system constraints, are gathered in
appendices.

%%%%%%%%%%%%%%%%%%%%%%%%%%%%%%%%%%%%%%%%%%%%%%%%%%%%%%%%%%%%%%%%%%%%%%%%%%%%%%
\section{Scalar-tensor theories of gravity}\label{sec2}

\subsection{Definition of the model}

In this article, we consider models in which gravity is described
by a scalar-tensor theory and in which the strength of the scalar
coupling may differ between CDM and ordinary matter. It follows
that in the Einstein frame the action takes the form
\begin{eqnarray}\label{actionEF}
 S &=& \int \frac{\dd^4x}{16\pi G_*}\sqrt{-g_*}\left[ R_*
        -2g_*^{\mu\nu} \partial_\mu\varphi_*\partial_\nu\varphi_*
        - 4V(\varphi_*)\right]\nonumber\\
   &&\qquad\qquad\qquad
   + S_V[A_V^2(\varphi_*)g^*_{\mu\nu};\psi_V]+ S_D[A_D^2(\varphi_*)g^*_{\mu\nu};\psi_{D}].
\end{eqnarray}
The action contains three arbitrary functions: the potential $V$,
the couplings to ordinary matter, $A_V$ ($V$ for visible sector),
and the coupling to CDM, $A_D$ ($D$ for dark sector). We will also
consider cases where both coupling functions are set equal to each
other. $G_*$ is the bare gravitational constant from which we
define $\kappa_*=8\pi G_*$.

The Jordan frame is typically defined as the frame for which the
metric is minimally coupled to matter. When $A_V=A_D$, there is a
unique definition of the Jordan frame. With two coupling
functions, we can define two different Jordan frames. Since
standard clocks and rods are made of standard matter, we define
the matter Jordan frame (MJF) to be that frame in which the
standard matter fields are minimally coupled to the metric. The
MJF metric will thus define length and time as measured by
laboratory apparatus so that all observations (time,
redshift,\ldots) have their standard interpretation in this frame.
This is also the frame in which the nuclear reaction rates take
their standard form, since they do not involve CDM. In this frame
the stress-energy tensor of visible matter is conserved. The
action then takes the form
\begin{eqnarray}\label{actionJF}
  S &=&\int \frac{\dd^4 x }{16\pi G_*}\sqrt{-g}
     \left[F(\varphi)R-g^{\mu\nu}Z(\varphi)\varphi_{,\mu}\varphi_{,\nu}
        - 2U(\varphi)\right]\nonumber\\
        &&   \qquad+ S_V[g_{\mu\nu};\psi_V]+ S_D[B^2(\varphi_*)g_{\mu\nu};\psi_{D}]
\end{eqnarray}
after performing the conformal transformation
\begin{equation}\label{jf_to_ef}
 g_{\mu\nu}^* = F(\varphi)g_{\mu\nu}
\end{equation}
with
\begin{eqnarray}
 \left(\frac{\dd\varphi_*}{\dd\varphi}\right)^2
              &=& \frac{3}{4}\left[\frac{\dd\ln F(\varphi)}{\dd\varphi}\right]^2
                  +\frac{Z(\varphi)}{2F(\varphi)}\label{jf_to_ef1}\\
 A_V(\varphi_*) &=& F^{-1/2}(\varphi)\label{jf_to_ef2}\\
 2V(\varphi_*)&=& U(\varphi) F^{-2}(\varphi)\label{jf_to_ef3}\\
 B(\varphi_*) &=& A_D/A_V.
\end{eqnarray}

We will denote all Einstein-frame quantities with a star ($*$). The
Einstein frame has the advantage of diagonalized the kinetic terms
for the spin-0 and spin-2 degrees of freedom so that the degrees of
freedom of the theory are more easily discussed. The strength of
the couplings of the scalar field to the matter and CDM fields is
characterized by
\begin{equation}\label{eqalpha}
 \alpha_i(\varphi_*)\equiv \frac{\dd\ln A_i}{\dd\varphi_*}
\end{equation}
and we also define
\begin{equation}\label{eqbeta}
 \beta_i(\varphi_*)\equiv \frac{\dd\alpha_i}{\dd\varphi_*}
\end{equation}
with $i=V,D$.

\subsection{Cosmological background equations}

\subsubsection{Space-time metrics}

We consider a Friedmann-Lema\^{\i}tre universe with metric in the
MJF
\begin{equation}
 \dd s^2 = -\dd t^2 + \scalefac^2(t)\gamma_{ij}\dd x^i \dd x^j
\end{equation}
where $\gamma_{ij}$ is the spatial me\-tric and $\scalefac$ the
scale factor.
In the Einstein frame, the metric takes the same form but with a
scale factor $\scalefac_*$ and a time coordinate, $t_*$ which are
related to their MJF counterparts by
\begin{equation}\label{JF2EF}
 R = A_V(\varphi_*) R_*,\qquad
 \dd t = A_V(\varphi_*)\dd t_*
\end{equation}
and the redshifts are related by
\begin{equation}\label{JF2EF2}
 1+z=\frac{A_{V0}}{A_V}(1+z_*).
\end{equation}

\subsubsection{Equations of motion in the Einstein frame}

The Friedmann equations in this frame take a form similar to those
of general relativity with a minimally coupled scalar field and a
fluid,
\begin{eqnarray}
 && 3\left(H_*^2 + \frac{K}{R_*^2}\right) =
   8\pi G_*\rho_* + \psi_*^2 + 2V(\varphi_*)\label{einsteinEF1}\\
 && - \frac{3}{R_*^2}\frac{\dd^2 R_*}{\dd t_*^2} = 4\pi G_* (\rho_* + 3P_*)
   + 2 \psi_*^2 - 2V(\varphi_*)\label{einsteinEF2}
\end{eqnarray}
where we have introduced $H_*=\dd \ln R_*/\dd t_*$ and
\begin{equation}\label{defpsistar}
 \psi_* = \dd\varphi_*/\dd t_*.
\end{equation}
$K$ is a positive, null or negative constant specifying the
curvature of the spatial sections. $\rho_*$ and $P_*$ are the
energy density and pressure of the total fluid.

In this model, the total energy can be split between the dark
sector ($\rho_D$) and the visible sector ($\rho_V$),
$$
 \rho = \rho_V + \rho_D.
$$
We assume that the dark sector component is described by a
constant equation of state $w_D$, which is set to $w_D=0$ for CDM.
The visible sector can be further decomposed between a
pressureless matter component $P_\mat=0$ and radiation
($P_\rad=\frac13\rho_\rad$), so that
$$
 \rho_V =\rho_\mat+\rho_\rad.
$$
We stress that $\alpha_\rad=\alpha_\mat=\alpha_V$ by construction,
in order to satisfy the universality of free fall. In the Einstein
frame, the evolution of these energy densities are dictated by the
matter conservation equation,
\begin{equation}\label{evEF}
 \frac{\dd\rho_{i*}}{\dd t_*} + 3H_*(\rho_{i*} + P_{i*}) =
 \alpha_i(\varphi_*)(\rho_{i*}-3P_{i*})\psi_*.
\end{equation}
The r.h.s. term proportional to $\psi_*$ is due to the scalar
interaction. In our case it reduces to $0$,
$\alpha_V\rho_{\mat*}\psi_*$ and $\alpha_D\rho_{D*}\psi_*$
for the radiation, standard matter and CDM.

The evolution of the scalar field is a Klein-Gordon equation
\begin{equation}\label{kgEF}
 \frac{\dd\psi_*}{\dd t_*} + 3H_*\psi_* = -\frac{\dd V}{\dd\varphi_*}
  -4\pi G_*\sum_i\alpha_i(\varphi_*)(\rho_{i*}-3P_{i*})\ ,
\end{equation}
where the sum is taken over $i=\rad,\mat,D$. The coupling term
reduces to
$$
 -4\pi G_*\sum_i\alpha_i(\varphi_*)(\rho_{i*}-3P_{i*}) =
 -4\pi G_*\left[\alpha_D\rho_{D*} + \alpha_V\rho_{\mat*}\right].
$$

\subsubsection{Implementing BBN}

The nuclear reaction network takes its standard form in the matter
Jordan frame, where the Lagrangian for the visible sector is not
affected by the existence of the scalar field so that the
cross-sections etc\ldots also take their standard form.

To compute the light elements abundances during BBN, one only
needs to know the expansion rate history, $H(z)$, from deep in the
radiation era up to today. It is thus convenient to express the
Hubble parameter in the MJF in terms of the one in
the Einstein frame, using Eq.~(\ref{JF2EF}), as
\begin{equation}\label{HHstar}
 A_V H = \left[H_* + \alpha_V(\varphi_*)\psi_*\right] .
\end{equation}
Eq.
(\ref{HHstar}) can also be expressed in the simple form
\begin{equation}\label{toto}
 A_VH = H_*\left[1 + \alpha_V(\varphi_*)\frac{\dd\varphi_*}{\dd
 p}\right]
\end{equation}
where $p$ is the number of e-folds of expansion in Einstein frame,
\begin{equation}\label{def_p}
 p = -\ln(1+z_*).
\end{equation}

Our numerical strategy follows exactly the one developed in
Ref.~\cite{couv}. We solve the Einstein equation in the Einstein
frame (see appendix~\ref{appB}) and then deduce $H(z)$ from
Eq.~(\ref{HHstar}) to compute the production of the light
elements.

\subsubsection{Evolution the energy densities}

The system of equations~(\ref{s1}-\ref{s6}) are expressed in terms of the energy densities of
the various components. Interestingly, these can be derived
analytically.
In the MJF, the densities are related to their
Einstein frame counterpart by
\begin{equation}\label{efmjf}
 \rho_* = A_V^4\rho,\qquad P_* = A_V^4 P.
\end{equation}
For matter in the visible sector, Eq.~(\ref{evEF}) takes the simple form
\begin{eqnarray}
 \dot\rho_V+3H(\rho_V+P_V) = 0
\end{eqnarray}
and is  trivially integrated to get
\begin{equation}
 \rho_{Vi} = \rho_{Vi0}(1+z)^{3(1+w_i)}\ ,
\end{equation}
for any component of the visible sector (i.e. $i=\rad,\mat$) with a
constant equation of state and where $z$ is the redshift defined
by $1+z=R_0/R$. Using Eqs.~(\ref{JF2EF}-\ref{JF2EF2}), we obtain
the evolution of the energy density in the  Einstein frame
\begin{equation}
 \rho_{{Vi}*} = \rho_{Vi0*}\left(\frac{A_{V}}{A_{V0}}\right)^{4-3(1+w_{i})}
 (1+z_*)^{3(1+w_{i})},
\end{equation}
where $\rho_{i0*}=A_{V0}^4\rho_{Vi0}$. We can also define the
associated density parameter by
\begin{equation}
 \Omega_{{Vi}0} \equiv \frac{8\pi G_*A_{V0}^2}{3H_0^2}\rho_{{Vi}0}.
\end{equation}

In the dark sector, the conservation equation
has a source term in the MJF,
\begin{eqnarray}
 \dot\rho_D+3H(\rho_D+P_D) = A_V^{-1}(\alpha_D-\alpha_V)(\rho_D-3P_D)\psi_*.
\end{eqnarray}
This equation can be obtained by plugging Eq.~(\ref{efmjf}) into
Eq.~(\ref{evEF}) for $i=D$ and using the relation~(\ref{JF2EF}) to
shift to derivative with respect to $t_*$. Interestingly, we can
find the solution of this equation analytically. In the CDM Jordan
frame, defined by $\tilde t$ and scale factor $\tilde R$,  CDM is
not coupled to the scalar field so that $\tilde\rho_D\propto\tilde
R^{-3(1+w_D)}$. Going back to the Einstein frame, this implies
that
$$
 \rho_{D*}\propto R_*^{-3(1+w_D)}A_D^{4-3(1+w_D)}
$$
and then in MJF,
\begin{equation}
 \rho_D = \rho_{D0}\left(\frac{A_{D}/A_V}{A_{D0}/A_{V0}}\right)^{4-3(1+w_D)}
 (1+z)^{3(1+w_D)}.
\end{equation}
Indeed, if the coupling of the scalar field is universal, the dark
matter and normal matter satisfy the same evolution equation.
In a general model where $A_V\not=A_D$, the two matter components
evolve differently. This is
related to the fact that both forms of matter do not experience the same
coupling to the scalar field.

We deduce that the ratio between the dark matter and baryonic
matter components is a priori not constant during the evolution of
the universe. It is given by
\begin{equation}
 \frac{\rho_{D*}}{\rho_{m*}}
        =\frac{\rho_D}{\rho_m}=
        \frac{\rho_{D0}}{\rho_{m0}}\frac{A_{V0}}{A_{D0}}\frac{A_D}{A_V}(\varphi_*)
   \equiv \Xi_0 \frac{B(\varphi_*)}{B_0}\ ,
\end{equation}
with $\Xi_0\equiv\rho_{D0}/\rho_{m0}$ and we define
$\tilde\Xi_0\equiv\Xi_0/B_0$. This is a main feature of this class
of models, which distinguishes it from universal scalar-tensor
theories.

Finally, the energy density of the scalar field in the Einstein
frame is given from Eq.~(\ref{einsteinEF1}) by
\begin{equation}\label{rhophiEF}
 \rho_{\varphi*} = \frac{1}{8\pi G_*}\left[\left(\frac{\dd\varphi_*}{\dd t_*}\right)^2
 + 2V(\varphi_*)\right]
\end{equation}
so that its energy density in the matter Jordan frame is
\begin{equation}\label{rhophiMJF}
 \rho_\varphi = A_V^{-4}\rho_{\varphi*}.
\end{equation}

%%%%%%%%%%%%%%%%%%%%%%%%%%%%%%%%%%%%%%%%%%%%%%%%%%%%%%%%%%%%%%%%%%%%%%%%%%%%%
\section{An Extended quintessence model}\label{sec3}

Before we move on to the most general type of models with two
coupling functions, we consider first an extension of the models
considered in Ref.~\cite{couv} with self-interaction potentials,
but with $A_V = A_D$. Our first examples are thus models of
extended quintessence in which the scalar field accounts for the
late time acceleration of the Universe while also being
responsible for a scalar interaction. Given the framework we have
set up in the previous section, we can now examine the
consequences on BBN of several choices of quintessence models.

As a first example, we consider an extended quintessence model of
the runaway type with a potential and coupling function given by
\begin{equation}\label{invpl}
 V=M^2\varphi_*^{-a},\qquad
 \ln A = C \varphi_*^{-b},
\end{equation}
where $a$, $b$ and $C$ are all positive and the mass scale $M \ll
\kappa^{-1/2}$ must be tuned so that $\Omega_\varphi \sim
\Omega_\Lambda$ today. It follows that
\begin{equation}
 \alpha = -Cb\varphi_*^{-b-1}.
\end{equation}

\begin{figure}[htb]
 \center\includegraphics[width=12cm]{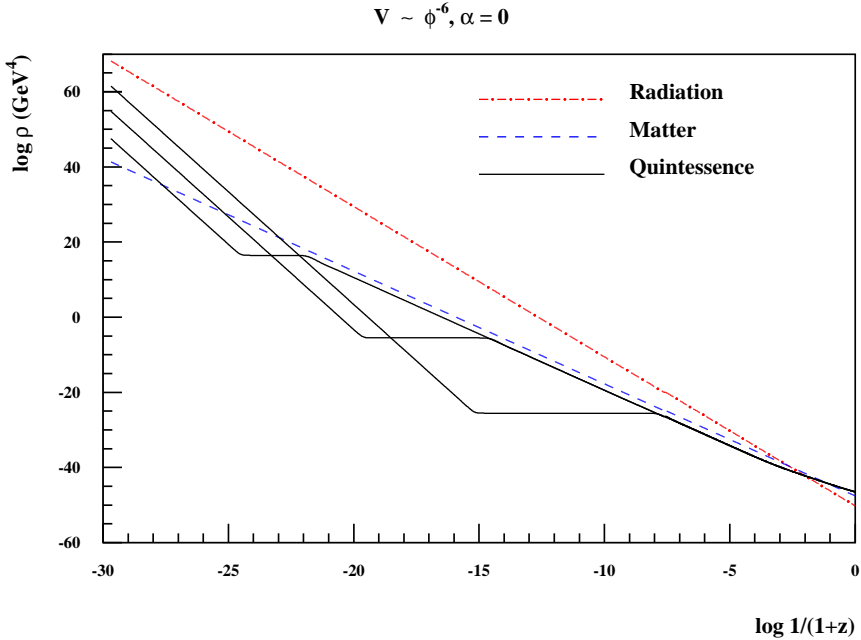}
 \caption{Evolution of the energy densities of the matter, radiation
 and scalar field as a function of the redshift for a model with
 potential~(\ref{invpl}) with $a=6$, $C=0$ and $M = 4\times10^{-17}$~s$^{-1}$  for different
 initial conditions, $\phi_{\rm in} = 2 \times 10^{-16}, 6 \times 10^{-16}$, and
 2 $10^{-15}$. The field initially enters a
 kinetic regime so that $\rho_\varphi\propto (1+z)^6$ and then reaches
 a slow-roll regime until its energy density is of the same order of magnitude
 as that of the matter and radiation. At this stage, it enters a tracking regime
 where its effective equation of state is $w_\varphi=0$ and $w_\varphi=-1/4$
 during the radiation and matter era respectively. It follows that the scalar
 field always ends up dominating the matter content of the universe at late times
 and gives rise to an accelerated phase of expansion.}
 % G4 : paw/BBN/couv08/Ria00/
 \label{figIPL}
\end{figure}

When $C=0$, this model is a standard quintessence model with an
inverse power law potential. The dynamics of this model has been
investigated in depth. In particular, there exist scaling
solutions with attractors of the dynamics of the scalar field
which behaves as a perfect fluid with an effective equation of
state
\begin{equation}
 w_\varphi = \frac{wa-2}{a+2},
\end{equation}
where $w$ is the equation of state of the fluid that dominates the
matter content of the universe. Figure~\ref{figIPL} depicts
the evolution of the energy densities as a function of the
redshift in this case and shows the quintessence attractor
mechanism typical for inverse power law quintessence potentials
when gravity is described by general relativity (i.e. $C=0$).

%%%%%%%%%%%%%%%%%%%%%%%%%%%%%%%%%%%%%%%%%%%%%%%%%%%%%%%%%%%%%%%%%%%%%%%%%%%%%%%
%%%%%%%%%%%%%%%%%%%%%%%%%%%%%%%%%%%%%%%%%%%%%%%%%%%%%%%%%%%%%%%%%%%%%%%%%%%%%%%

When $C\not=0$, and as discussed in detail in Ref.~\cite{couv},
we expect two main effects. First during BBN, the field dynamics
is modified during electron-positron annihilation and then the
late time dynamics is modified because of the coupling to
non-relativistic matter. Let us investigate them by assuming that
$V=0$.

As detailed in \S~III.A.2 of Ref.~\cite{couv}, the dynamics during
electron-positron annihilation is governed by the equation
$$
 \frac{2}{3-\varphi_*^{\prime2}}\varphi_*''+ \frac{2}{3}\varphi_*' =
  -\alpha(\varphi_*)\Sigma_e,
$$
where $\Sigma_e\equiv(\rho_e-3P_e)/\rho_\rad$ and prime denotes a
derivative with respect to $p$. In a non-kinetic regime, this
equation can be approximated by
$$
  \varphi_*''+ \varphi_*' =
  \frac{3}{2}Cb\varphi_*^{-b-1}\Sigma_e.
$$
As in the case of a quadratic coupling~\cite{couv}, the field is
frozen to some initial value $\varphi_{\xin*}$ during the
radiation era \cite{condil} prior to the electron-positron annihilation. Then
the source term $\Sigma_e$ acts temporarily inducing the evolution
of $\varphi_*$ which then settles to another constant value
$\varphi_{\xout*}$ in the radiation era. Because $\ln A$ does not have a
minimum, the field will shift towards larger values so that
$\varphi_{\xout*}>\varphi_{\xin*}$.

During the matter era, the evolution equation reduces to
$$
 \frac{2}{3-\varphi_*^{\prime2}}\varphi_*''+ \varphi_*' =
  -\alpha(\varphi_*)=Cb\varphi_*^{-b-1},
$$
Since the minimum of the coupling function is at infinity, it follows
that the scalar field is simply attracted toward this value and
does not undergo damped oscillations as in the case of a quadratic
coupling. In particular, in the slow-roll regime, it behaves as
$$
 \varphi_*=\varphi_{\rm eq*}\left[1+ (2+b)\frac{Cb}{\varphi_{\rm eq*}^{2+b}}
 (p-p_{\rm eq}) \right]^{\frac{1}{2+b}},
$$
where $p_{\rm eq}$ is the value of $p$ at the matter-radiation
equality.

\begin{figure}[htb]
\center{\includegraphics[width=8cm]{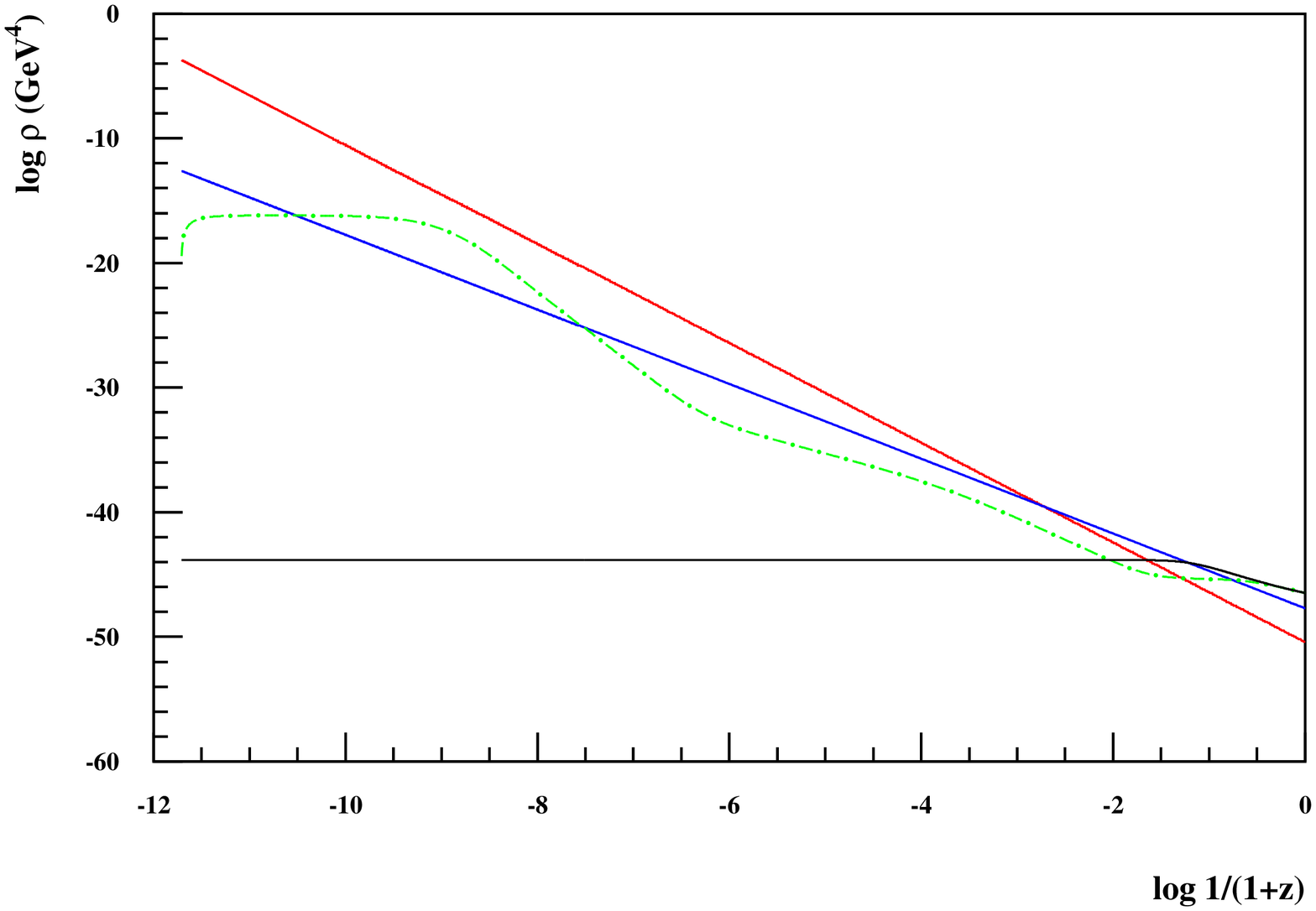}
         \includegraphics[width=8cm]{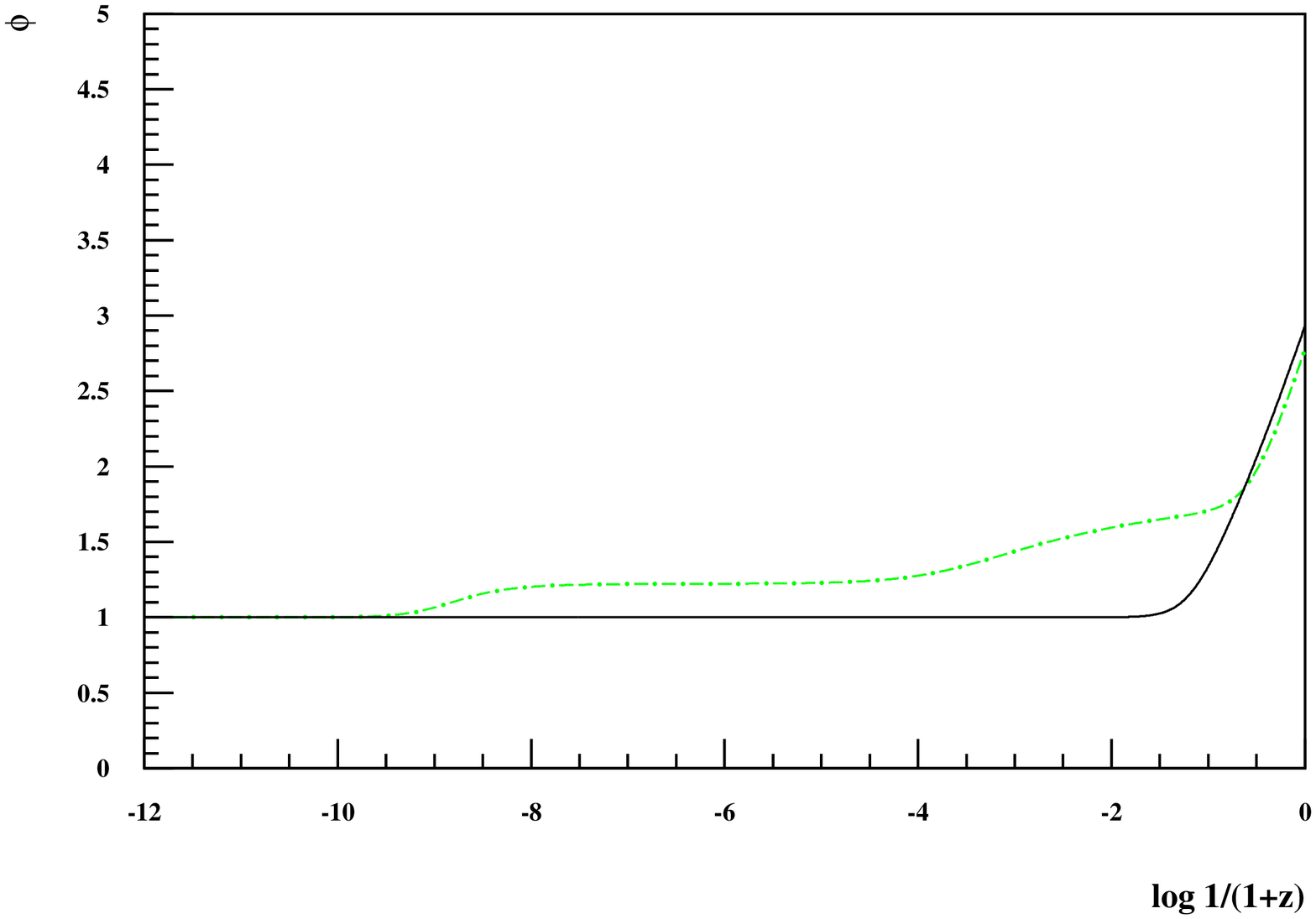}}
 \caption{(Left): Evolution of the energy densities of the matter (blue line), radiation (red line)
 and scalar field as a function of the redshift for a model with
 potential~(\ref{invpl}) with $\phi(z=10^{12})=1$,  $a=b=6$ and
 $C=0$ (solid black line) or $C=0.2$ (dash dotted green line).
 (Right): Evolution of $\phi$ for the same conditions.}
 \label{figIPLA}
\end{figure}

In the general case where both the potential and the coupling are
effective, the solution is first attracted toward the quintessence
scaling solution during the radiation era during which the
coupling is not efficient. The various mass thresholds further
drive $\varphi_*$ toward larger values so that it reaches the
quintessence tracking solution more rapidly. Then, during the
matter era, the evolution is driven by the potential and coupling
which both drive $\varphi_*$ to infinity so that the scalar-tensor
theory is attracted toward general relativity.
Figure~\ref{figIPLA} shows the magnitude of these effects.

%%%%%%%%%%%%%%%%%%%%%%%%%%%%%%%%%%%%%%%%%%%%%%%%%%%%%%%%%%%%%%%%%%%%%%%%%%%%%%%
%%%%%%%%%%%%%%%%%%%%%%%%%%%%%%%%%%%%%%%%%%%%%%%%%%%%%%%%%%%%%%%%%%%%%%%%%%%%%%%

Quintessence fields coupled to the Maxwell term, $F^2$, in the
action, will induce a variation of the fine structure constant.
There are a few models~\cite{other,lop} which can simultaneously
account for the purported variation in quasar absorption
systems~\cite{varalpha} and the  Oklo and meteoritic
constraints~\cite{Oklo,OPQ,fuj}. Here we consider the model of
Ref.~\cite{lop} with the potential
\begin{equation}\label{lopV}
 V(\varphi_*) = V_0
 \exp\left(\frac{\lambda}{2}\varphi_*^2\right),
\end{equation}
which is representative of a class
of models which possess a minimum in the self-interaction potential and is
closely related to the one studied in Refs.~\cite{pot1,pot2}.
The potential is normalized by taking $V_0$ to correspond to the present
vacuum energy density. The
coupling function in this model is also specified and related to the potential
\begin{equation}\label{lopA}
 A(\varphi_*) =\left[\frac{b+V(\varphi_*)/V_0}{1+b}\right]^n
\end{equation}
where $1+b>0$. Its main feature is to have a common minimum with
that of the potential, and allows the system to recover general
relativity when the scalar field is at the minimum. This is in the
spirit of the least coupling principle~\cite{dn} whose goal is to
suppress the scalar interaction when the field has reached its
minimum. Note that in Ref.~\cite{lop} where the scalar field was
also responsible for a variation of the fine structure constant,
constraints were derived such that $n$ needed to be very small. In
the current context,  the field is universally coupled so that it
is responsible only for a variation of the gravitational constant,
which is much less constrained than the time variation of the fine
structure constant~\cite{jpuct,ctes2}.

\begin{figure}[htb]
 \center{\includegraphics[width=8cm]{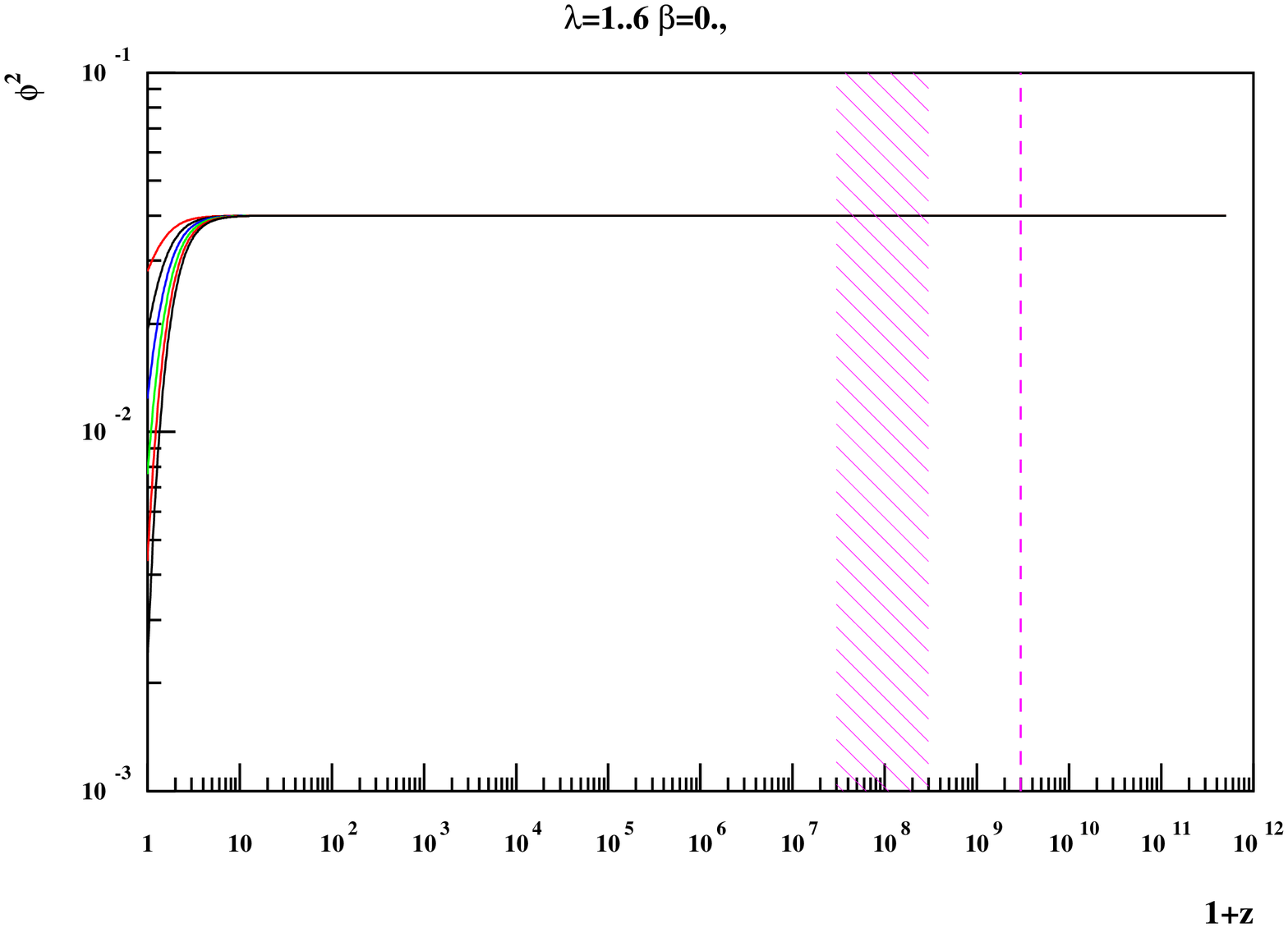}
         \includegraphics[width=8cm]{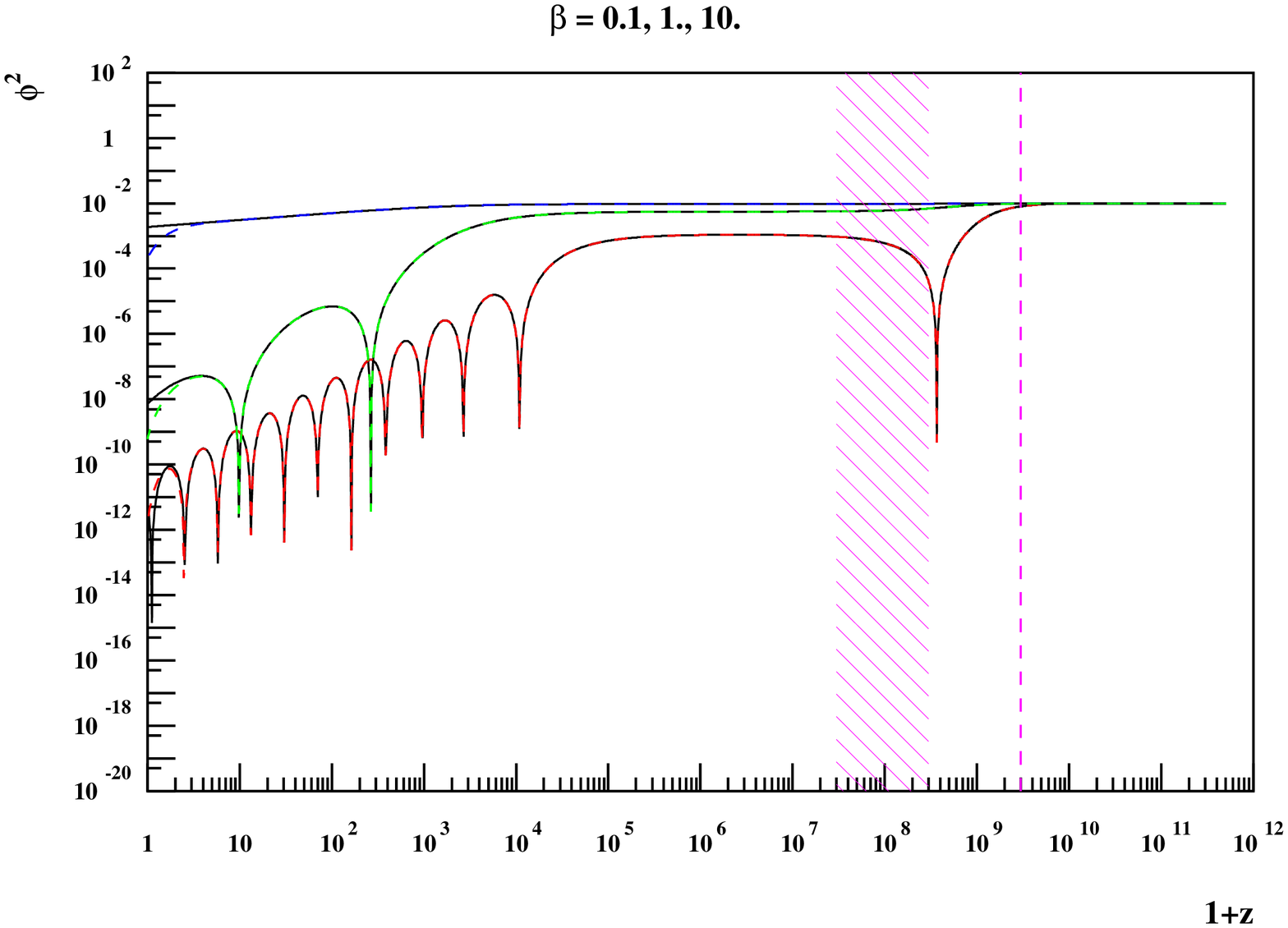}}
 \caption{Evolution of $\varphi_*$ as a function of the
 redshift for models with potential and coupling defined in
 Eqs.~(\ref{lopV}-\ref{lopA}). (Left): we set $\beta$=0 and
 $\lambda$ = 1, 2, 3, 4, 5 and 6, so that we are dealing with a
 quintessence model and no modification of general relativity.
 (Right): we set $\beta$= 0.1, 1. and 10. without a potential
 (solid) or with a potential and $\lambda$ = 6 (dashed). Differences
 with and without a potential appear
 only at the very late times.
 The vertical dashed line corresponds to the time of $n/p$
freeze-out and the shaded region to the epoch of BBN.}
 % G4 : paw/BBN/couv08/Lee04/
 % G4 : paw/BBN/couv08/Lee04/
\label{lop}
\end{figure}

For small values of the field $\varphi_*$, $A$ can be approximated
by $1 + n \lambda \varphi_*^2/2(1+b)$, so that
we can write
\begin{equation}\label{lopalphabeta}
\alpha \approx \frac{n \lambda \varphi_*}{1+b} \qquad \beta \approx \frac{n\lambda}{1+b}
\end{equation}
For $\beta = 0$, we show the evolution of $\varphi_*$ as a function of redshift in
Figure \ref{lop}a. We see that for most of the evolution $\varphi_*$ is constant
largely due to the very small value of $V_0$.  At late times,
the field evolves toward the minimum at the origin.  This type of model
(with very small $\beta$) was constructed to account for the possible evolution of the
fine structure constant.  In panel b), we show the corresponding evolution when
$\beta$ takes values 0.1, 1, and 10. We see that
as $\beta$ is increased, the evolution of $\varphi_*$ begins at higher redshift.
When $\beta = 1 (10)$, we also see that the field undergoes a few (many) oscillations
about the origin.

%%%%%%%%%%%%%%%%%%%%%%%%%%%%%%%%%%%%%%%%%%%%%%%%%%%%%%%%%%%%%%%%%%%%%%%%%%%%%%%
%%%%%%%%%%%%%%%%%%%%%%%%%%%%%%%%%%%%%%%%%%%%%%%%%%%%%%%%%%%%%%%%%%%%%%%%%%%%%%%

%%%%%%%%%%%%%%%%%%%%%%%%%%%%%%%%%%%%%%%%%%%%%%%%%%%%%%%%%%%%%%%%%%%%%%%%%%%%%
\section{Non-universal massless dilaton with quadratic
couplings}\label{sec4}

We now consider  models in which $A_V\not=A_D$. For simplicity we
focus on models with a constant potential and we assume that the
two coupling functions are quadratic, which gives us a natural
extension of the model studied in Ref.~\cite{couv}. We set
\begin{equation}
 A_V=\exp\left(\frac{1}{2}\beta_V\varphi_*^2\right),\quad
 A_D=\exp\left(\frac{1}{2}\beta_D\varphi_*^2\right),\quad
 V=V_0,
\end{equation}
and we define
\begin{equation}
 a_i(\varphi_*)=\ln[A_i(\varphi_*)].
\end{equation}
This is also a generalization of Ref.~\cite{dgg} which was
restricted to Brans-Dicke models.

$V_0$ is constant so that the scalar degree of freedom remains
massless. As in the previous model considered [Eq.~(\ref{lopV}] it
is associated with a constant energy density in the Einstein frame
\begin{equation}
 \rho_{\Lambda*} \equiv  V_0/4\pi G_*\ ,
\end{equation}
but scales as $A_V^{-4}$ in the matter Jordan frame. It follows
that the total energy density is $\rho_T=\rho+\rho_\Lambda$. Note
that the couplings are such that this component is not coupled to
$\varphi_*$, that is
\begin{equation}
 A_\Lambda=1\ ,\qquad
 \alpha_\Lambda=0.
\end{equation}
We note that the tests on the deviations from general
relativity in the Solar system involve only $\alpha_V$ and
$\beta_V$ since they are performed with ordinary matter.

\subsection{Dynamics}

\subsubsection{Reduced equation}

The dynamics of this non-universal system can be discussed using
$p$ as a time variable. Since $\dd p=H_*\dd t_*$, we can deduce
that $\psi_* = H_*\varphi_*'$ and we can rewrite the Friedmann
equation~(\ref{s3}) as
\begin{equation}
 (3-\varphi_*^{\prime2})H_*^2 = 8\pi G_*(\rho_*+\rho_{\Lambda*}).
\end{equation}
The derivative of Eq.~(\ref{s2}) with respect to $p$ implies that
$$
 \varphi_*''+\left(\frac{H_*'}{H_*}+3\right)\varphi_*' =
 -\frac{4\pi G_*}{H_*^2}\sum_i\alpha_i(1-3w_i)\rho_{i*}.
$$
Now, it can be checked that
$$
 \frac{H_*'}{H_*}=-\frac{1}{2}'(3-\varphi_*^{\prime2})
 (1+\tilde w)\varphi'_*-\varphi_*^{\prime2},
$$
where $\tilde w$ includes the contribution of the cosmological
constant,
\begin{equation}
 \rho_*(1+w)=(\rho_*+\rho_{\Lambda*})(1+\tilde w)\ .
\end{equation}
Thus, we conclude that the dynamics is described by
\begin{eqnarray}\label{edyn}
 \frac{2}{3-\varphi^{\prime2}}\varphi_*''+(1-\tilde w)\varphi_*' &=&
 -\sum_i\alpha_i(1-3w_i)\frac{\rho_{i*}}{\rho_*+\rho_{\Lambda*}} \\
 &=&
  -\alpha_V\frac{\rho_{\mat*}}{\rho_*+\rho_{\Lambda*}}-\alpha_D(1-3w_D)
  \frac{\rho_{D*}}{\rho_*+\rho_{\Lambda*}}
\end{eqnarray}
This generalizes the equation used in Refs.~\cite{dn,couv,dp} to
the case of two couplings. We emphasize that this equation holds
only when $V=V_0$ is constant and for $K=0$.

It is interesting to note that if the cosmological constant
is negligible and if the dark matter and the baryonic matter have
the same equation of state $w_D=w$ then Eq.~(\ref{edyn}) is
equivalent to the one that would be obtained with one single fluid
with equation of state $w$ with a universal scalar-tensor theory
with an effective coupling function
\begin{equation}
 A_{\rm eff}(\varphi_*) = A_V(\varphi_*) + \tilde \Xi_0 A_D(\varphi_*),
\end{equation}
which depends on the ratio of dark matter and baryonic matter, as
well as initial conditions, through $\Xi_0$.

\subsubsection{Phase space}

Neglecting the cosmological constant, which dominates only in the
last e-fold, Eq.~(\ref{edyn}) reduces to
\begin{eqnarray}\label{edynsimplofoed}
 \frac{2}{3-\varphi_*^{\prime2}}\varphi_*''+(1- w)\varphi_*' &=&
  -\frac{\beta_V + \beta_D\Xi_0\hbox{e}^{\frac12(\beta_D-\beta_V)
  (\varphi_*^2-\varphi_{*0}^2)}}{1+\Xi_0\hbox{e}^{\frac12(\beta_D-\beta_V)
  (\varphi_*^2-\varphi_{*0}^2)}+\frac{\rho_\rad}{\rho_\mat}}\varphi_*.
\end{eqnarray}
Let us focus on the matter dominated era. Eq.
(\ref{edynsimplofoed}) indicates that the attraction toward
general relativity will depend crucially on the relative signs of
$\beta_V$ and $\beta_D$ and on the form of the effective coupling
function $A_{\rm eff}$ which enters on the r.h.s. of the Klein-Gordon
equation.

We have the following different possibilities:
\begin{itemize}
 \item $\beta_V=0$: the scalar-tensor theory is pure general
 relativity in the visible sector. Only cosmology can set
 constraints on $\beta_D$ while all Solar system constraints are
 always satisfied.
 \item $\beta_D=0$: the theory is attracted toward general
 relativity only if $\beta_V>0$.
 \item $\beta_D>0$ and $\beta_V>0$: $a_V$ and $a_D$ have the same
 minimum at $\varphi_*=0$ so that the effective function $a_{\rm eff}$
 also has a unique minimum in $\varphi_*=0$ which is an attractor
 of the dynamics. The scalar-tensor theory is thus
 attracted towards general relativity, both for the matter and the
 dark sectors, during electron-positron annihilation and
 the late time matter era when $\varphi_*\rightarrow0$.
 \item $\beta_D<0$ and $\beta_V<0$: Neither $a_V$ nor $a_D$ have a
 minimum and thus $a_{\rm eff}$ does not have a minimum. $\varphi_*$ runs to infinity
 when either matter or dark matter is dominant. The scalar-tensor theory
 thus drifts away from general relativity. We shall thus discard
 this case since local constraints can be satisfied only
 at the price of an extreme fine-tuning of the initial conditions.
 \item $\beta_D\beta_V<0$: This situation is more complex.
 $\varphi_*=0$ is always an extremum of $a_{\rm eff}$ but not necessary a
 minimum. Two conditions have to be considered.

 First, if
 \begin{equation}\label{cond1}
 \Phi\equiv
 \frac{2}{\beta_D-\beta_V}\ln\left(-\frac{\beta_V}{\beta_D \tilde \Xi_0}\right)<0,
 \end{equation}
 $a_{\rm eff}$ has a {\em unique} extremum at $\varphi_*=0$ otherwise it
 has two others given by
 \begin{equation}
  \varphi^2_{\rm att *}=\varphi_{0*}^2
   + \frac{2}{\beta_D-\beta_V}\ln
    \left(-\frac{\beta_V}{\beta_D}\frac{\rho_{\mat0}}{\rho_{D0}}\right).
 \end{equation}
 Second, if
 \begin{equation}\label{cond2}
 \tilde\beta\equiv\beta_V+ \tilde \Xi_0\beta_D>0,
 \end{equation}
 the extremum at $\varphi_*=0$ is a minimum.

 The signs of these two
 functions set the shape of the function $a_{\rm eff}$ which
 determines the late time attractor of the Klein-Gordon equation.
Note that $\Phi$ changes sign when $\tilde\beta$ does.
 Fig.~\ref{fig:pspsace1} summarizes the various possibilities in
 the phase space while
 Fig.~\ref{fig:poteff} depicts the modification to $a_{\rm eff}$
 when the values of $\beta_D$ and $\beta_V$ vary and when the
 signs of $\Phi$ and $\tilde\beta$ change.

 \begin{figure}[htb]
 \center{\includegraphics[width=9cm]{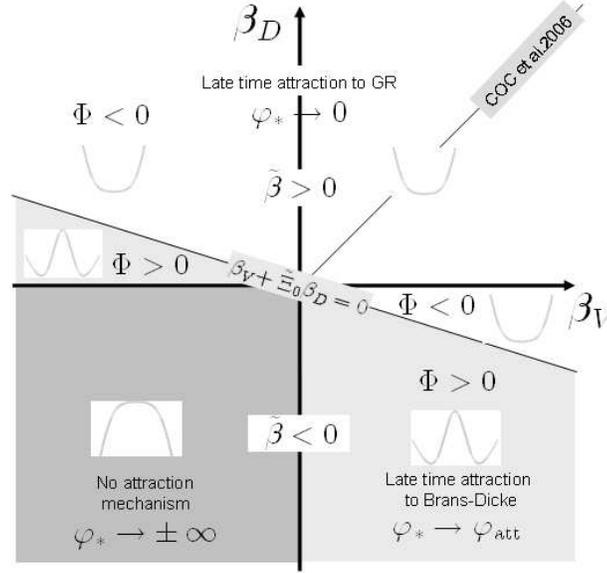}}
 \caption{The $(\beta_V,\beta_D)$ parameter space with the signs of
 $\Phi$ and $\tilde\beta$ [defined in Eqs.~(\ref{cond1}) and ~(\ref{cond2})]
 which determine the shape of $a_{\rm eff}$ and thus
 the late time attractors of the Klein-Gordon equation.
  Three
 possibilities appear: late time attraction toward general relativity for $\tilde\beta>0$,
 late time attraction toward a Brans-Dicke theory when $\tilde\beta<0$ and $\beta_D\beta_V<0$
 and runaway when $\beta_D<0$ and $\beta_V<0$. The field is respectively
 attracted toward 0, $\pm\varphi_{\rm att}$ and $\pm\infty$.}
 \label{fig:pspsace1}
\end{figure}

\begin{figure}[htb]
 \center{\includegraphics[width=8cm]{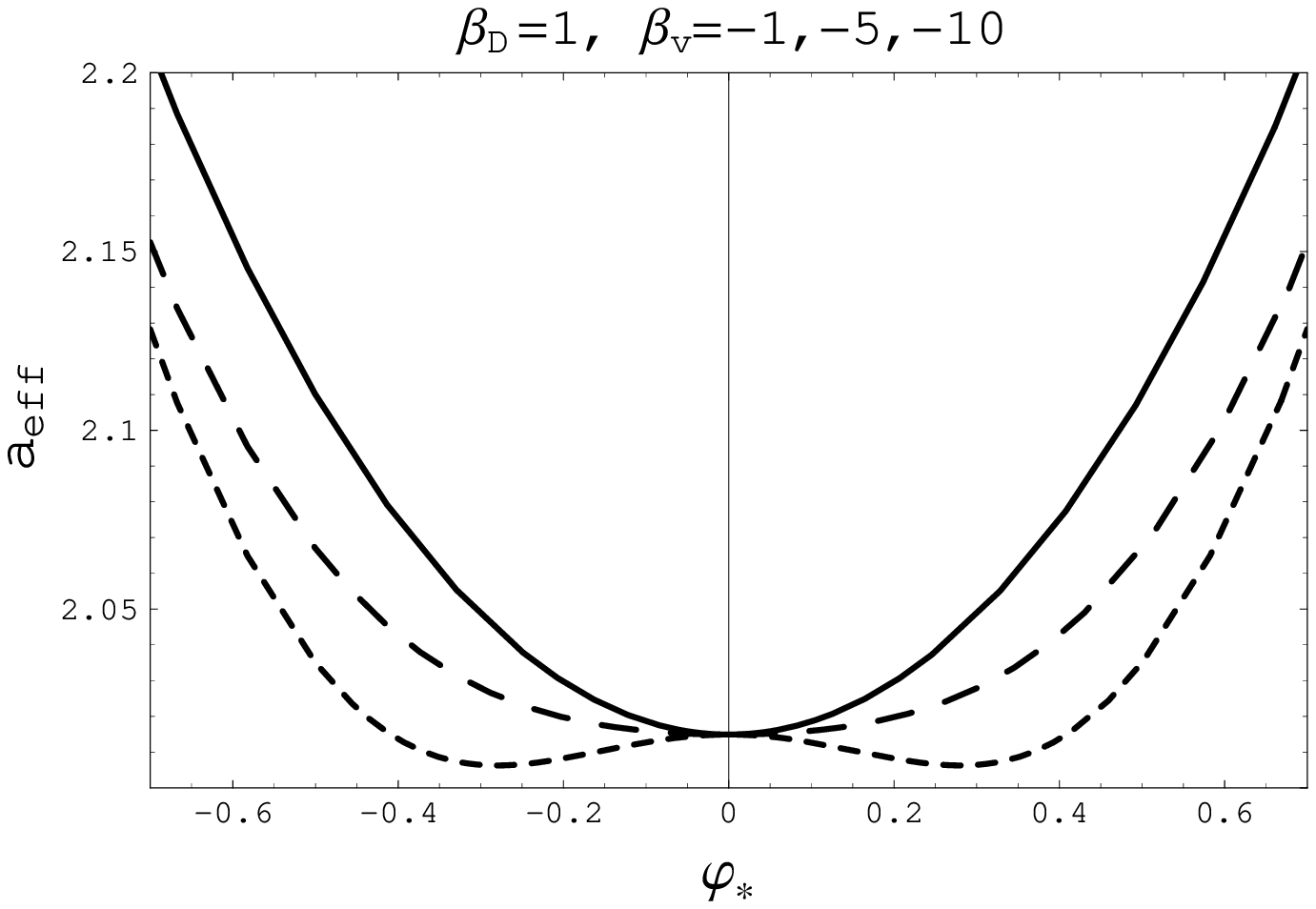}\includegraphics[width=8cm]{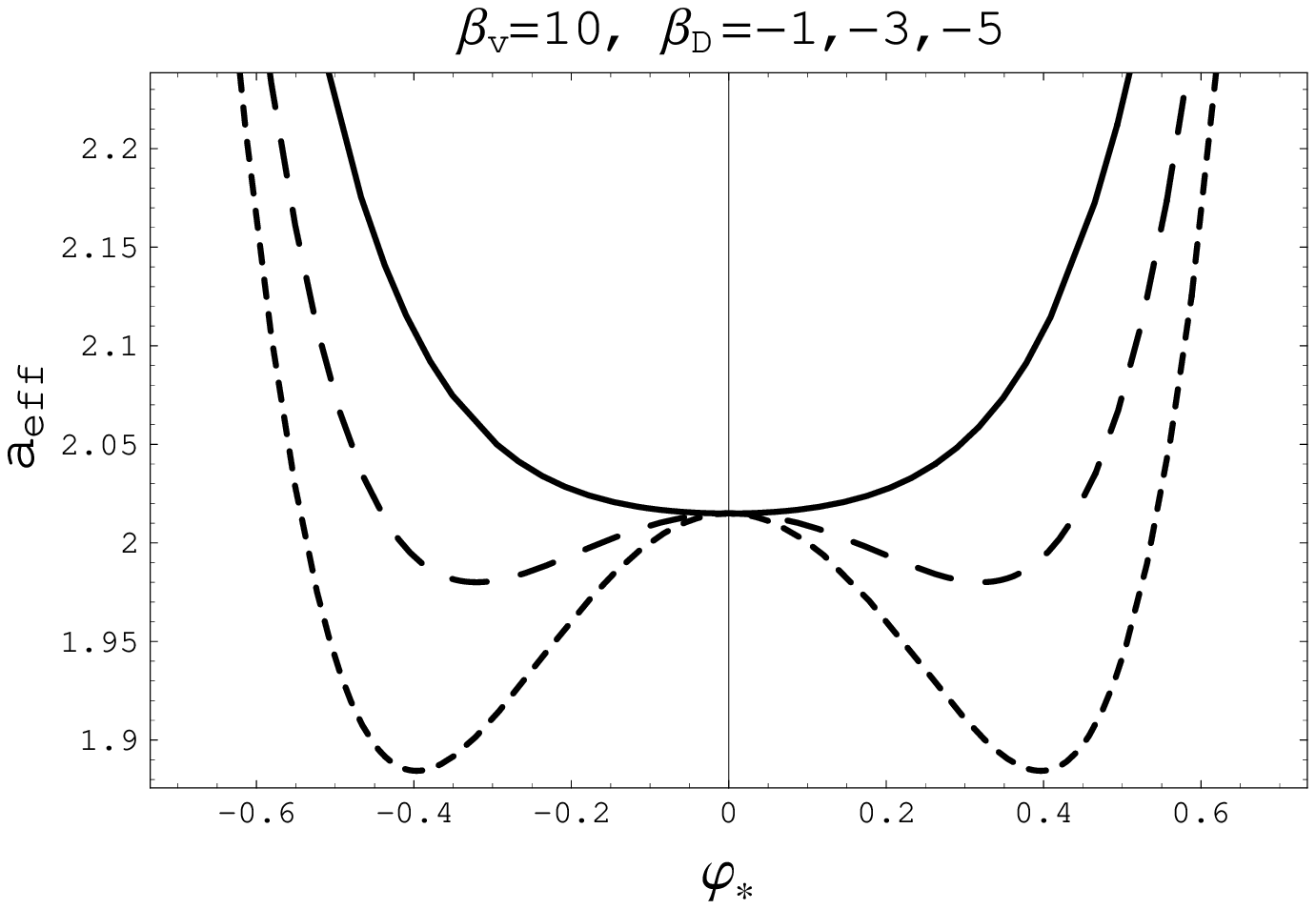}}
 \caption{Modification of the shape of $a_{\rm eff}$
 with the values of $\beta_D$ and $\beta_V$ and with the
 signs of $\Phi$ and $\tilde\beta$.}
 \label{fig:poteff}
\end{figure}

 Given the shape of $a_{\rm eff}$, we have two possibilities for the
 late time dynamics:
 \begin{itemize}
 \item $\varphi_*=0$ is the only minimum. The Klein-Gordon
 equation involves $A_{\rm eff}$ so that $\varphi_*$ is attracted
 toward 0 during the matter
 era. Since this corresponds to the minimum of $a_V$ and $a_D$,
 the scalar-tensor is also attracted toward general relativity.
 \item $\varphi_*=0$ is a local maximum. In this case $\varphi_*$ is attracted
 toward $\varphi_{\rm att}\not=0$. This does not coincide with a minimum
 of $a_V$ and $\alpha_V\rightarrow\beta_V\varphi_{\rm att}$. The
 scalar-tensor is thus attracted towards a Brans-Dicke theory, that
 is a scalar tensor theory with $\ln A= \lambda\varphi_*$.
 Fig.~\ref{fig:attBD} illustrates such a case where $\varphi_*$
 cannot relax to the minimum of the coupling function $A_V$ because
 the dynamics is dictated by $A_{\rm eff}$. It follows from
 Eq.~(\ref{cond1}) that
 $$
 \alpha_V^2\rightarrow\alpha_{V0}^2 +
 2\frac{\beta_V}{\beta_D-\beta_V}\ln\left(-\frac{\beta_V}{\beta_D\Xi_0}\right).
 $$
According to the value of the parameters, the asymptotic
Brans-Dicke parameter can be compatible with the observational
constraints. Indeed, the theory can be temporarily attracted
toward general relativity due to the mass thresholds in the
primordial universe if $\beta_V>0$. In this case, $\alpha_{V0}$
can be low enough to pass Solar system tests.

  \begin{figure}[htb]
 \center{\includegraphics[width=8cm]{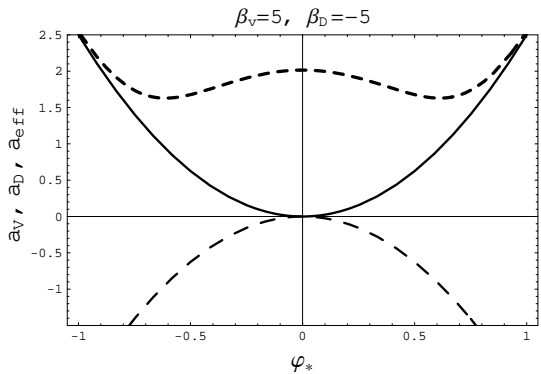}}
 \caption{When the minimum of $a_{\rm eff}$ (dashed line) differs from the one of $a_V$
 (solid line) and
 $a_D$ (long dashed line), the dynamics of the scalar field drives $\varphi_*$ to a minimum
 of $A_{\rm eff}$ so that the scalar-tensor theory is attracted toward
 a Brans-Dicke theory.}
 \label{fig:attBD}
\end{figure}
 \end{itemize}
\end{itemize}

\begin{center}
\begin{figure*}[ht]
\unitlength=1cm
 \thicklines
\begin{picture}(14,12)
 \put(-1,0) {\framebox(16,12){}}
 \put(-1,6){{\psfig{file=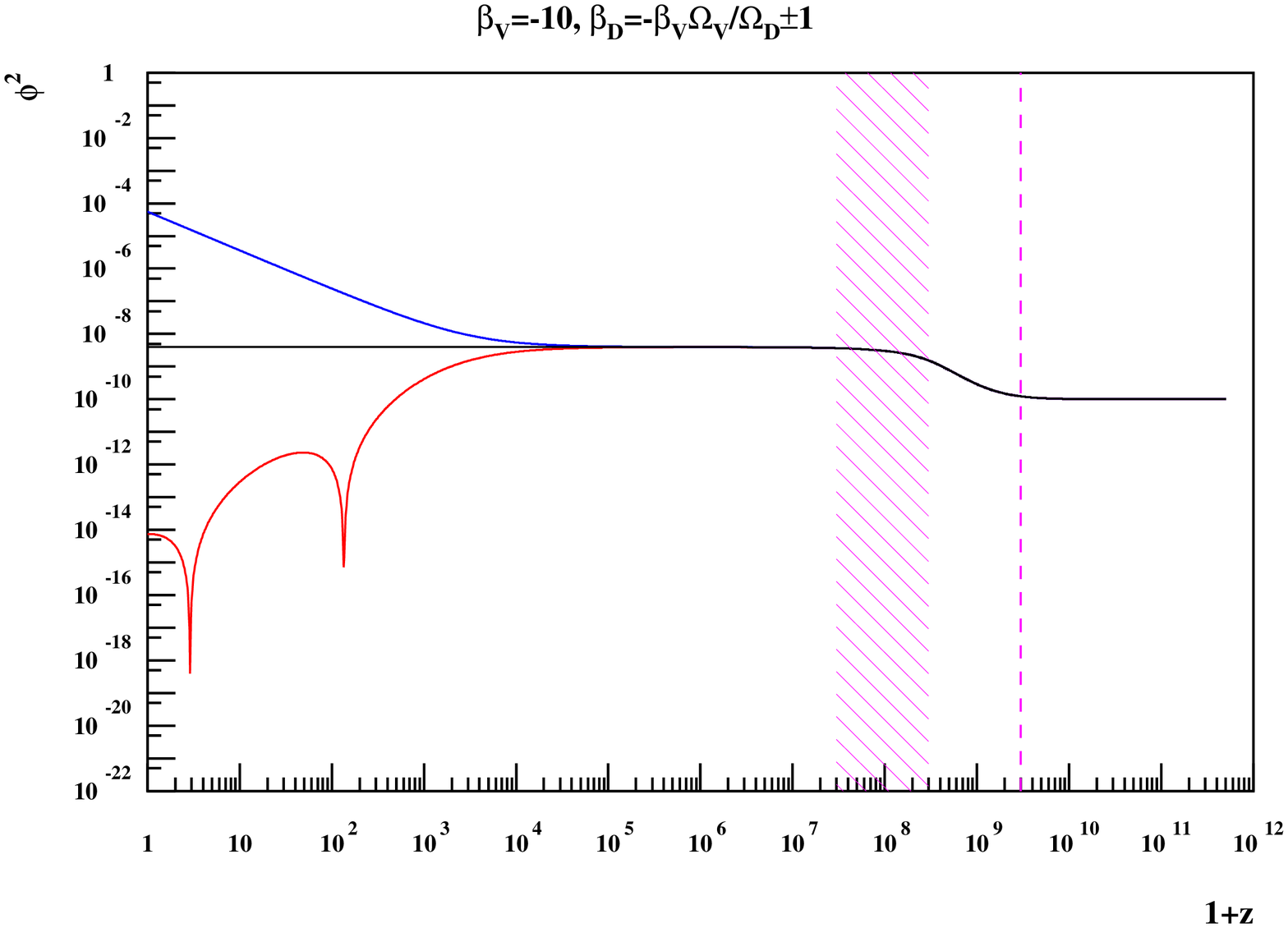,width=8cm}}}
 \put(7,6){{\psfig{file=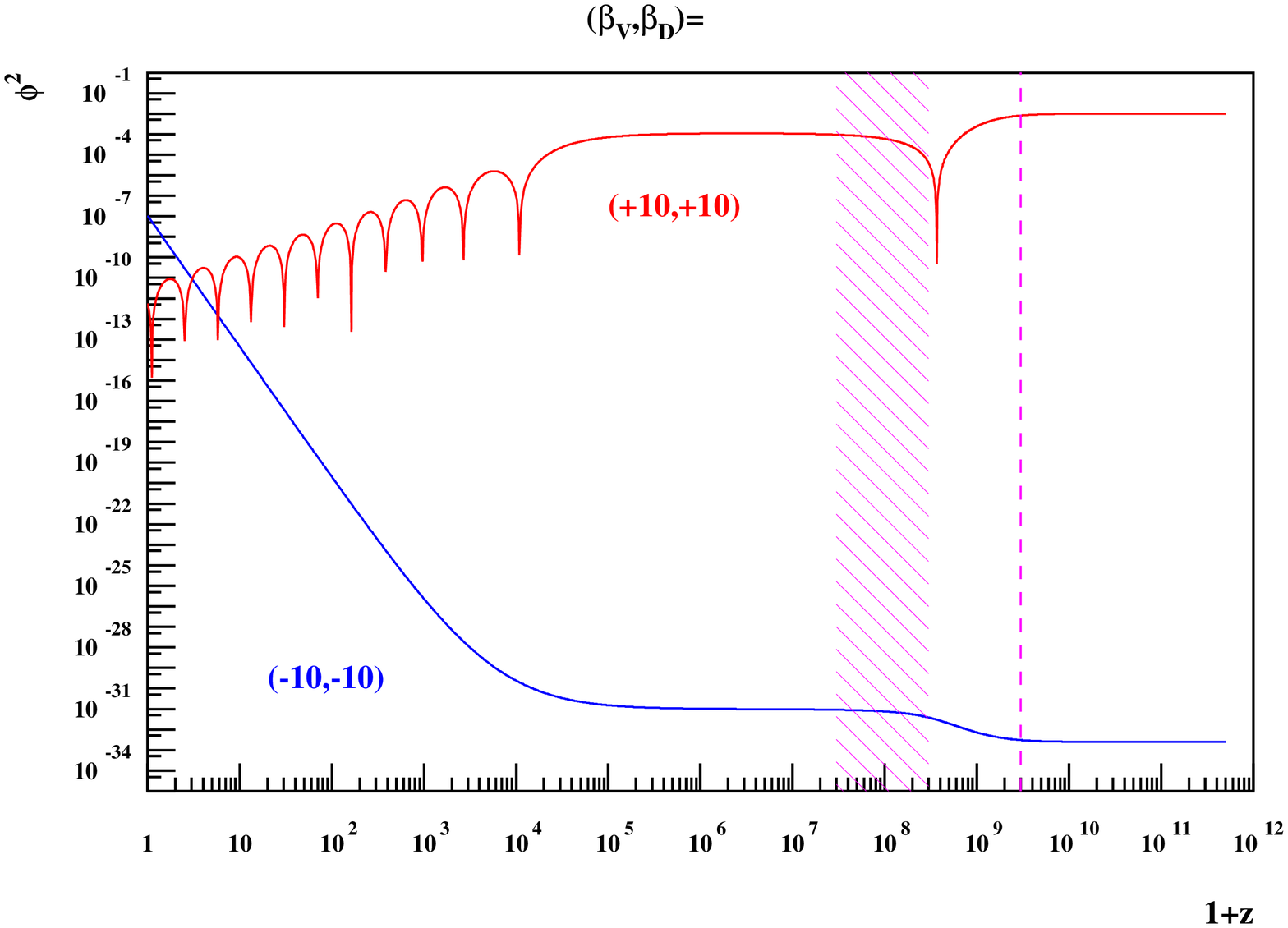,width=8cm}}}
 \put(7,0){{\psfig{file=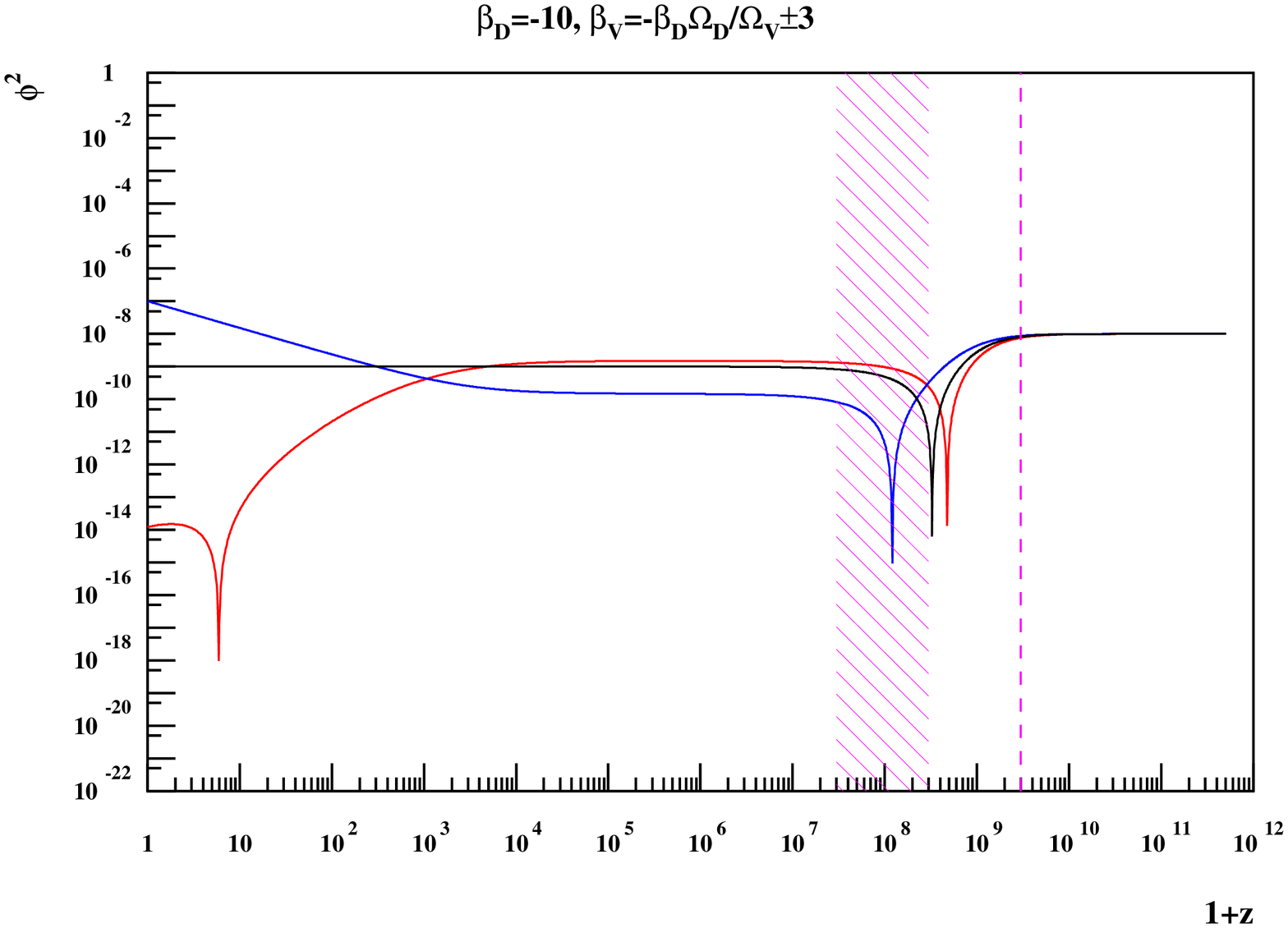,width=8cm}}}
 \put(-.5,.8){{\psfig{file=fig4a.eps,width=4.5cm}}}
 \put(5.0,4.0){\vector(1,1){2.7}}
 \put(3.35,4.0){\line(1,0){1.65}}
 \put(1,4.7){\vector(0,1){1.8}}
 \put(3.5,1.75){\vector(1,0){3.9}}
\end{picture}
\caption{(Lower left): Phase space of the solutions.
 (Upper-left) when $\beta_V<0$ the theory is dragged away from
 general relativity during BBN and then, depending on the sign of
 $\tilde\beta$, attracted towards or away from general relativity at
 late times. The field is at the minimum of the coupling
 function today only if $\tilde\beta=0$ (as seen by the horizontal line at late times).
 (Lower-right) when $\beta_D<0$ the theory is attracted
 toward general relativity during BBN and then, according to the sign of
 $\tilde\beta$ attracted or not toward general relativity at
 late time.
 (Upper-right) Both $\beta$s are positive and the theory is always
 attracted toward general relativity. The opposite is true when both
 $\beta$s are negative.
 } \label{figmethod}
\end{figure*}
\end{center}

As discussed above, Figures~\ref{fig:pspsace1} and~\ref{figmethod} summarize the
field dynamics. Three possibilities appear: late time attraction
towards general relativity for $\tilde\beta>0$, late time
attraction toward a Brans-Dicke theory when $\tilde\beta<0$ and
$\beta_D\beta_V<0$ and runaway when $\beta_D<0$ and $\beta_V<0$.
In each of these cases, the field is attracted
towards 0, $\pm\varphi_{\rm att}$ and $\pm\infty$, respectively.
Only models
such that the effective parameter $\tilde\beta$ is positive, enjoy
a late time attraction toward general relativity.

Fig.~\ref{figmethod} connects the phase space
$(\beta_V,\beta_D)$ with specific examples of model evolution
which highlight the different possibilities. Fig. \ref{fig:pspsace1} has
been miniaturized in the lower left of Fig.~\ref{figmethod}
and with evolutionary examples for each quadrant except that
for which $\beta_D,\beta_V < 0$ and the theory exhibits runaway
behavior.

%------------------------------------------------------------
%--------------------------------------------------------------

In the upper left of Fig.~\ref{figmethod}, we show the evolution of $\varphi_*^2$
as a function of redshift for $\tilde \beta = -1, 0, +1$, all with $\beta_V = -10$.
If $\beta_V$ is negative the theory is dragged away from
general relativity.  This is seen by the increase in $\varphi_*^2$
prior to BBN. This could be compensated at late times
if $\tilde\beta>0$ as seen in the lower curve with $\tilde \beta = 1$
which undergoes two oscillations about the minimum.
Note that when $\tilde\beta=0$ the Klein-Gordon equation has no
source term even after the end of the radiation era so that
$\varphi_*$ remains frozen to the same value than during the
radiation era. The scalar-tensor theory can be attracted toward
general relativity only due to the effects of the mass thresholds.
Therefore no evolution is seen after BBN when $\tilde \beta = 0$.
When $\tilde \beta = - 1$, the field continues to evolve
away from $\varphi_*^2 = 0$ when matter domination sets in.
However, and as discussed in detail
in Ref.~\cite{couv}, there are many other mass thresholds prior to
electron-positron annihilation and a $\beta_V<0$ model
would have to be fine-tuned to be attracted close enough to
general relativity today. We shall thus restrict our analysis to
$\beta_V>0$.

Models with $\beta_V>0$ are attracted toward general relativity
during electron-positron annihilation (see discussion below).
Then, if $\beta_D<0$, the theory is either attracted toward
general relativity or a Brans-Dicke theory depending on the sign
of $\tilde\beta$. As the analysis above also shows, in the latter
case the theory is attracted toward a value of $\varphi_*$
different from $\varphi_*= 0$. Examples of  models with
$\beta_V > 0$, but $\beta_D < 0$ and $\tilde \beta = -3, 0, +3$ are
shown in the lower right plot of
Fig.~\ref{figmethod}. Each case moves towards the origin (and undergoes one
oscillation about the origin) prior to and during BBN.  At late times, we
see again that the model with $\tilde \beta > 0$ continues to evolve towards the
origin after matter domination, while model with $\tilde \beta < 0$ evolves
away from general relativity. As before the case with $\tilde \beta = 0$ shows
no further evolution.

Finally, in the upper right panel of Fig.~\ref{figmethod}, we show
an example of a model with $\beta_D = \beta_V = +10$. The field
moves through the origin once during electron-positron
annihilation, and then later continues to oscillate about the
origin.  For completeness, we also show an example with $\beta_D =
\beta_V = -10$ in this panel which shows the run away behavior at
late times.

In conclusion, we will concentrate our analysis on the case in
which both $\beta_V$ and $\beta_D$ are positive (see upper-right
plot of Fig.~\ref{figmethod}). This corresponds to models that are
attracted toward general relativity without involving a tuning of
the different parameters and the initial conditions for the value
of $\tilde\beta$ to be positive.

\subsubsection{Different regimes}

Deep in the radiation era, the cosmological constant the dark
matter and baryonic matter components are negligible so that
Eq.~(\ref{edyn}) simplifies to
\begin{eqnarray}
 \frac{2}{3-\varphi^{\prime2}}\varphi_*''+\frac{2}{3}\varphi_*' &=& 0.
\end{eqnarray}
As a consequence,  the field is frozen at
a constant value and the initial conditions can be chosen as
$$
\varphi'_{*\xin}=0\ , \qquad \varphi_{*}=\varphi_{*\xin}= {\rm
constant}\ .
$$

Still in the radiation era, when the universe cools below the mass
of some species, $\chi$, this species becomes non-relativistic and
induces a non-vanishing contribution to the r.h.s. of
Eq.~(\ref{edyn}). The most important mass threshold for the BBN
predictions is the last of these thresholds, the one associated
with electron-positron annihilation. Previous mass thresholds are also
important and drive the scalar-tensor theory toward general
relativity (see \S~III.A.3 of Ref.~\cite{couv}). We assume that
the dark sector is not affected by any mass thresholds after muon
annihilations.  In that case, we can neglect the effect of the
dark sector and we end up with the same result as in
Ref.~\cite{couv},
\begin{eqnarray}\label{kgqq1}
 \frac{2}{3-\varphi_*^{'2}}\varphi_*''
 +\frac{2}{3}\varphi_*' + \Sigma_e(T)\beta_V\varphi_*=0\ ,
\end{eqnarray}
with
\begin{eqnarray}
 \Sigma_e(T) &=& \frac{15}{\pi^4}\frac{g_e}{g_*(T)}z_e^2\int_{z_e}^\infty
 \frac{\sqrt{x^2-z_e^2}}{\hbox{e}^x+1}\dd x\ .
\end{eqnarray}
This implies that we are driven toward general relativity during
the electron-positron annihilation only if $\beta_V>0$ (see right
half of Fig.~\ref{figmethod}).

After electron-positron annihilation, the universe is still
dominated by radiation and the scalar field freezes at a
constant value, $\varphi_{*\xout}$.
BBN can place a constraint on the value of $a_\xout = a(\varphi_{*\xout})$. Unfortunately,
this constraint depends on $a_\xin$ which is unknown. To compare
these constraints to those obtained in the Solar system, we
need to relate $a_\xout$ to $a_0$. Thus, our code integrates the
evolution equation up to the present, so that we obtain $a_0$
directly. For the particular case of a vanishing potential or as
long as the field is slow rolling, $\varphi'\ll3$,
Eq.~(\ref{edyn}) takes the slightly simplified form
\begin{equation}
\frac{2}{3}\varphi_*''+(1- w)\varphi_*' =
  -\frac{\beta_V + \beta_D\Xi_0\hbox{e}^{\frac12(\beta_D-\beta_V)
  (\varphi_*^2-\varphi_{*0}^2)}}{1+\Xi_0\hbox{e}^{\frac12(\beta_D-\beta_V)
  (\varphi_*^2-\varphi_{*0}^2)}}\varphi_*.
\end{equation}
Contrary to the case of a universal coupling (\S~III.5 of
Ref.~\cite{couv}) this equation cannot be integrated analytically
in the matter-radiation era.

The field is attracted toward its
minimum and at late times we can assume that $\varphi_*\ll1$, in which
case the equation of evolution reduces to
$$
 \frac{2}{3}\varphi_*''+(1- w)\varphi_*' +
 \omega^2\varphi_*=0,\qquad
 \omega^2\equiv\frac{\beta_V +
 \beta_D\tilde\Xi_0}{1+\tilde\Xi_0}
$$
It follows that the field evolves as
\begin{eqnarray}
 \varphi_*(p)&=& \hbox{e}^{-\frac34 p}\left[ A\cos\left(\frac34 rp\right) +
               B\sin\left(\frac34 rp\right)\right],\label{ephisol}\\
             &&  \hbox{e}^{-\frac34 p}\left[ A\cosh\left(\frac34 rp\right) +
               B\sinh\left(\frac34 rp\right)\right],
\label{scaling}
\end{eqnarray}
respectively if $\omega^2<\frac83$ and $\omega^2>\frac83$ with
$$
 r\equiv \sqrt{\left|1-\frac83 \omega^2\right|}.
$$
This allows us to compute the period of the last oscillations,
\begin{equation}\label{pulssol}
 \Delta p = \frac83\frac{\pi}{r(\beta_V,\beta_D,\tilde\Xi_0)}.
\end{equation}
Unfortunately we can not compute the phase, which is
required in order to match to the solution in the radiation era.

\begin{figure}[htb]
 \center\includegraphics[width=8cm]{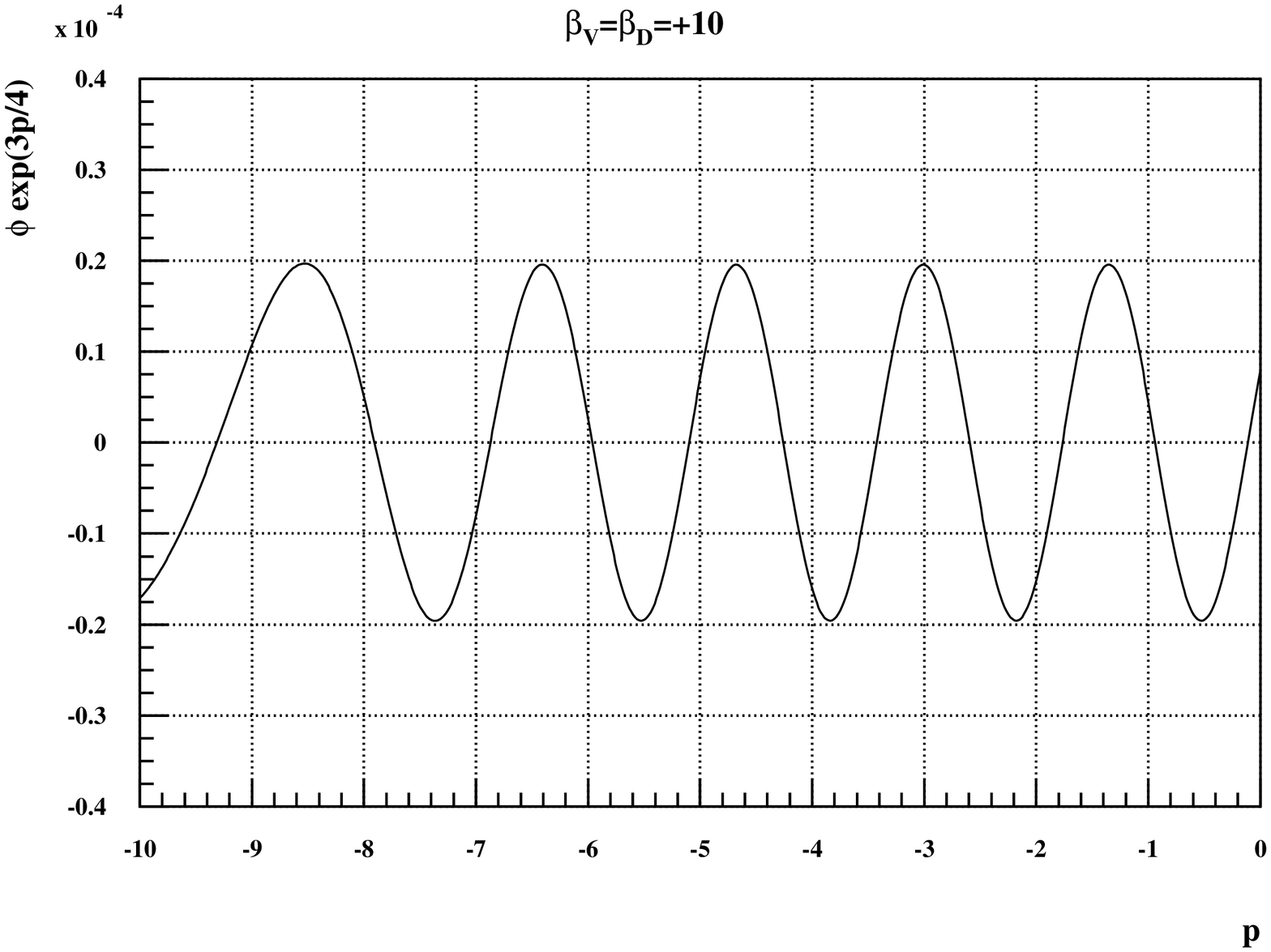}
        \includegraphics[width=8cm]{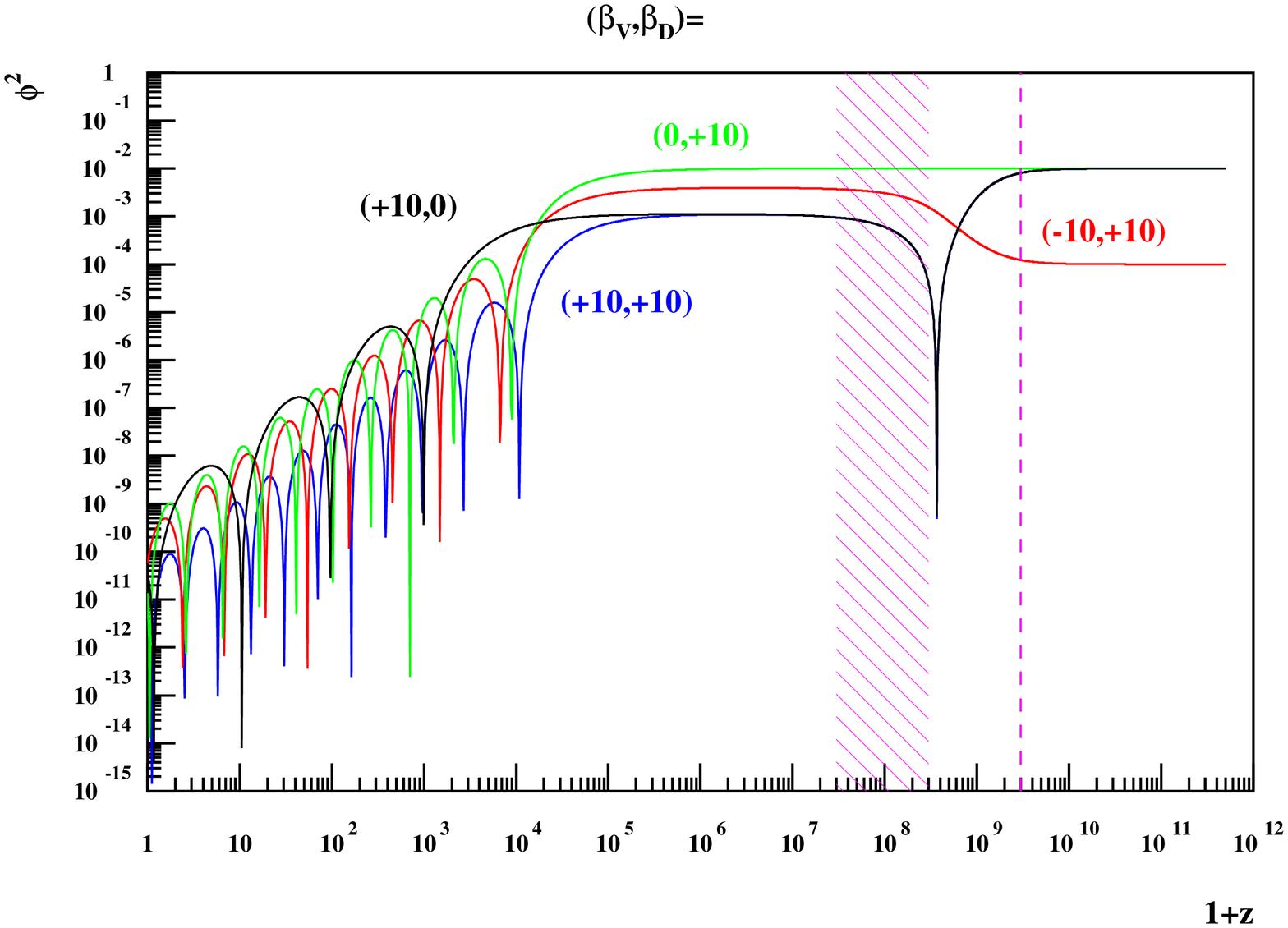}
 \caption{(Left): Evolution of $\varphi_*\exp(\frac34p)$ as a
 function of redshift. The frequency of the oscillations depends
 on $\beta_V$, $\beta_D$ and the energy densities as in
 Eq.~(\ref{pulssol}).
          (Right):  Some solutions with $\tilde\beta>0$
 summarizing the effect of $\beta_V$ during BBN and the evolution
 during the matter era.
 } \label{figbevo}
\end{figure}

An example of such an evolution is depicted in the left panel of
Fig.~\ref{figbevo} with $\beta_V = \beta_D = +10$ as in the upper
right of Fig.~\ref{figmethod}. Here we see the same oscillations
scaled by $e^{3p/2}$ as expected from Eq.~(\ref{scaling}) with
period given by Eq.~(\ref{pulssol}). In the right panel of
Fig.~\ref{figbevo}, we show several cases (as labelled) showing
the late time scaling and oscillations. We can see that the field
remains frozen during the radiation era up to the kick during
electron-positron annihilation. When the universe starts to be
matter dominated the field undergoes damped oscillations with a
period given by Eq.~(\ref{pulssol}).

%%%%%%%%%%%%%%%%%%%%%%%%%%%%%%%%%%%%%%%%%%%%%%%%%%%%%%%%%%%%%%%%%%%%%%%%%%%%%%%
%%%%%%%%%%%%%%%%%%%%%%%%%%%%%%%%%%%%%%%%%%%%%%%%%%%%%%%%%%%%%%%%%%%%%%%%%%%%%%%

\subsection{Constraints}

\subsubsection{BBN}

\begin{figure}[htb]
 \center\includegraphics[width=5cm]{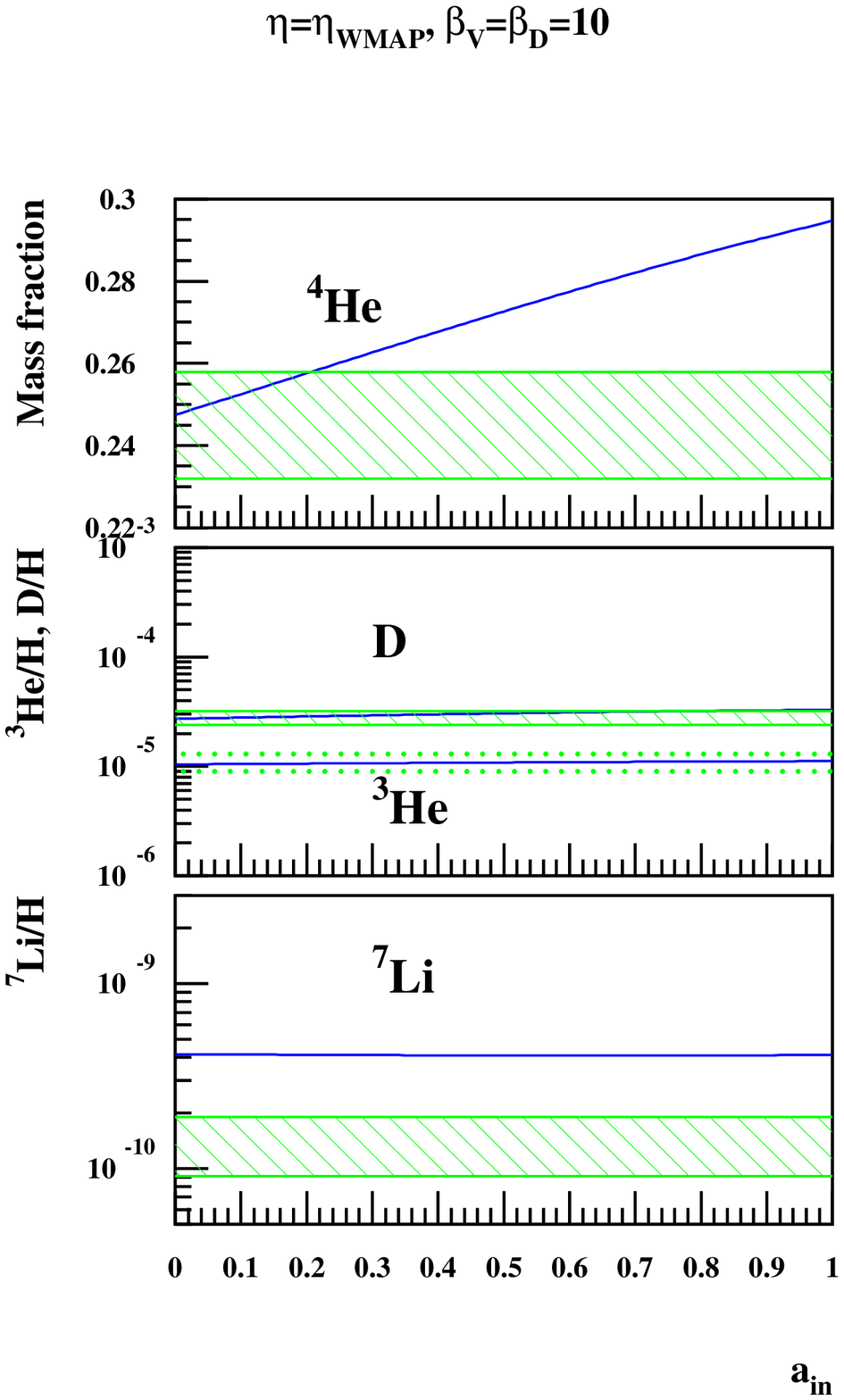}
 \includegraphics[width=5cm]{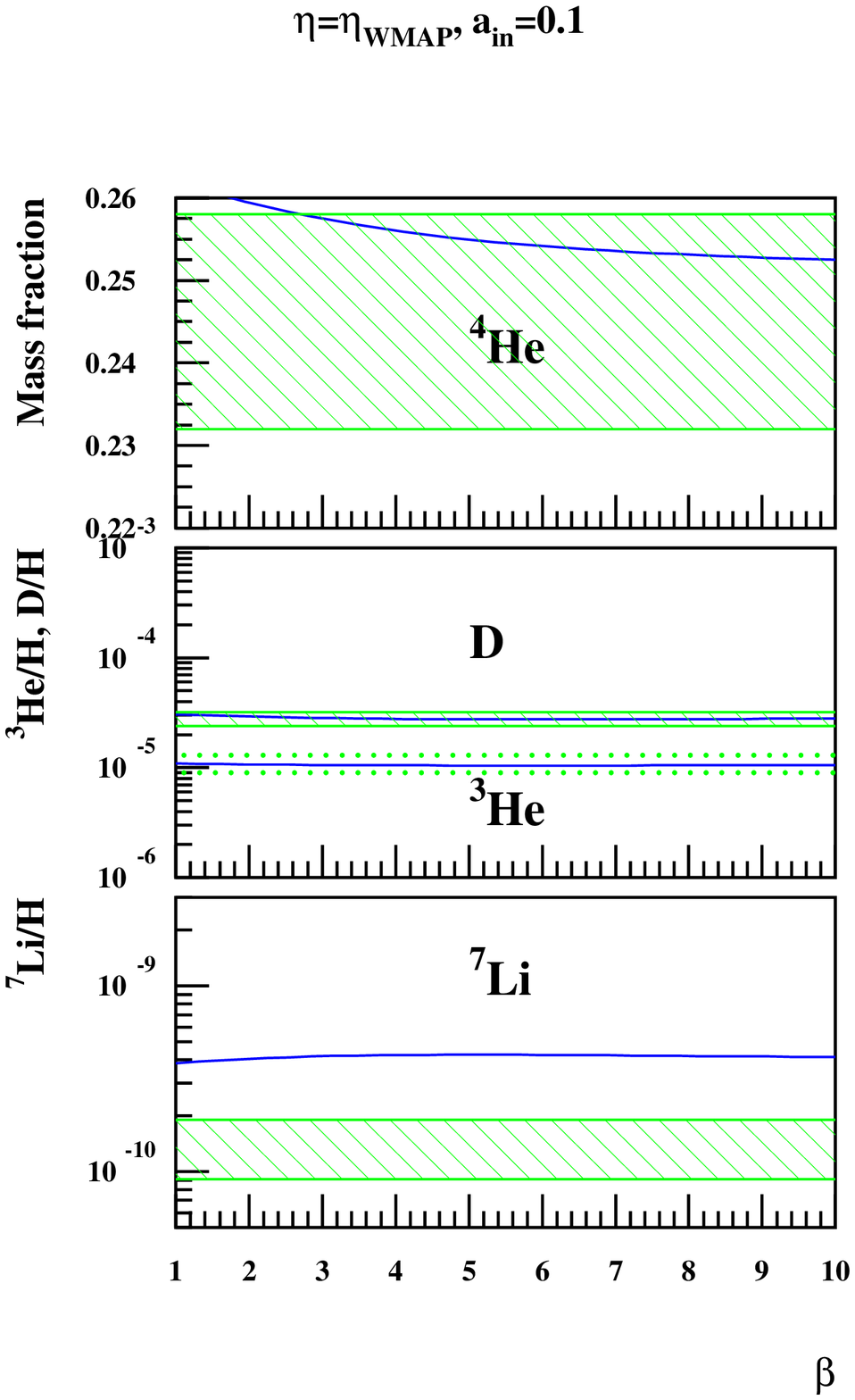}
 \includegraphics[width=5cm]{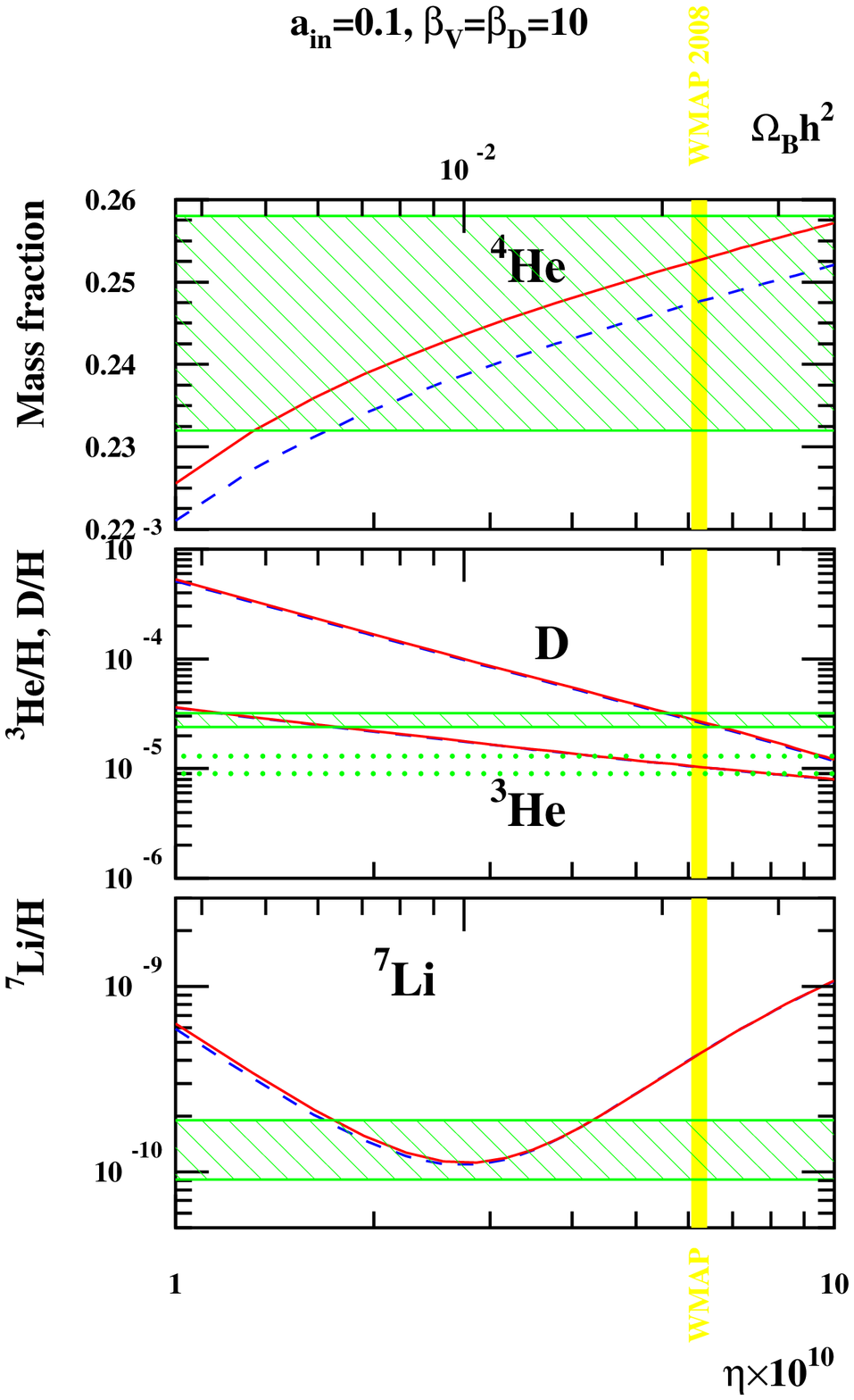}
 \caption{(Left): Light element abundance as a function of
 $a_\xin$ when $\eta=\eta_{\rm WMAP}$ and $\beta_V=10$.
 (Middle): Same as a function of $\beta_V$ when $\eta=\eta_{\rm WMAP}$
 and $a_\xin=0.1$. (Right): Same as a function of $\eta$ for
 $\beta_V=10$ and $a_\xin=0.1$. The standard model is shown by the lower
 (blue) dashed curves.
}
 \label{fig:Yp1}
\end{figure}

\begin{figure}[htb]
 \center\includegraphics[width=10cm]{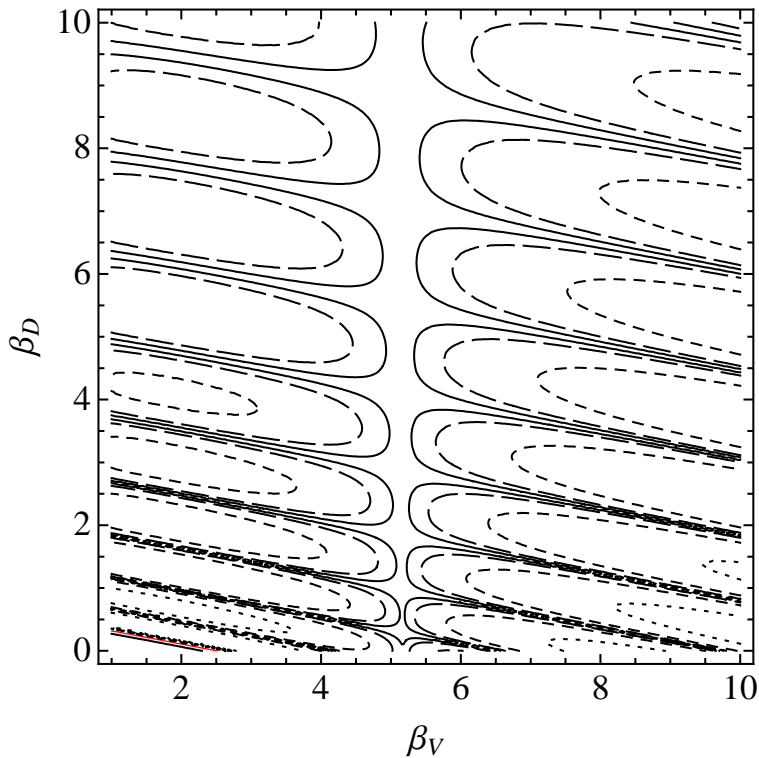}
 \caption{Constraints on $\alpha_{V0}$ in the plane
 $(\beta_V,\beta_D)$, respectively for $\log\alpha_{V0}=-5$
 (solid line), $-4.5$ (long dashed lines) $-4$ (dashed lines)
 and $-3.5$ (dotted lines). The red line in the lower left corner shows
 the solar system constraint of $\log\alpha_{V0}=-2.5$}
 \label{figVD}
\end{figure}

We proceed as in Ref.~\cite{couv} and vary the baryon to photon
ratio, $\eta$, the values of $\beta_V$ and $\beta_D$ as well as
the initial conditions $a_\xin$ but we assume that the other
cosmological parameters are fixed to their standard values. We
compute the light element abundances as a function of these four
parameters. Fig.~\ref{fig:Yp1} shows the effect of the independent
variation of each of these parameters. This allows one to set the
BBN constraints on these parameters which are then propagated
until today to infer constraints on $\alpha_{V0},\beta_V$ and
$\beta_D$. Fig.~\ref{figVD} summarizes these constraints on the
$(\beta_V,\beta_D)$ plane obtained from BBN.

%%%%%%%%%%%%%%%%%%%%%%%%%%%%%%%%%%%%%%%%%%%%%%%%%%%%%%%%%%%%%%%%%%%%%
%%%%%%%%%%%%%%%%%%%%%%%%%%%%%%%%%%%%%%%%%%%%%%%%%%%%%%%%%%%%%%%%%%%%%

In the left panel of Fig.~\ref{fig:Yp1}, we show the resulting
light element abundances as a function of the parameter $a_\xin$
for fixed values of $\beta_V=\beta_D = 10$ and $\eta$ fixed at the
WMAP value of $6.23 \times 10^{-10}$ \cite{dunkley}. As one can
see in this panel and others to the right, there is very little
dependence of \he3, D, or \li7 on either $a_\xin$ or $\beta_V$.
Note that the \li7 abundance is always in excess of the
observations for this value of $\eta$ (see e.g. Ref.~\cite{cfo5}).
In contrast, there is a relatively strong dependence of the \he4
abundance with $a_\xin$. The shaded regions correspond to the
range of the observational determinations as described in Appendix
C. The upper limit on the \he4 abundance of 0.259 places a
constraint on $a_\xin \le 0.2$, when $\beta_V=10$. In the middle
panel of Fig.~\ref{fig:Yp1}, we show the element abundances as a
function of $\beta_V$ for fixed $a_\xin = 0.1$ and the WMAP value
for $\eta$. At this value of $a_\xin$, we find a lower bound on
$\beta_V > 2.8$. Finally, in the right panel of
Fig.~\ref{fig:Yp1}, we show the abundances as a function of $\eta$
for fixed $a_\xin = 0.1$ and $\beta_V = \beta_D = 10$. This choice
of parameters leads to an increase in the \he4 abundance of
roughly 0.005 over the standard model with $a_\xin = 0$ and
$\beta_V = 0$ as shown by the lower set of blue dashed curves.

%%%%%%%%%%%%%%%%%%%%%%%%%%%%%%%%%%%%%%%%%%%%%%%%%%%%%%%%%%%%%%%%%%%%%
%%%%%%%%%%%%%%%%%%%%%%%%%%%%%%%%%%%%%%%%%%%%%%%%%%%%%%%%%%%%%%%%%%%%%

Now, for each value of the pair $(\beta_V,\beta_D)$, we can compute
the abundances of \he4, \he3, D and \li7 as a
function of $a_\xin$ which allows us to constrain the values
of this parameter from the observational data. We
have set $\eta=\eta_{\rm WMAP}$, though in principle, one could
derive combined constraints on the pair of inputs ($a_\xin, \eta$).
One can check that this is a safe approximation and
that the deviations from general relativity for the models
considered do not lead to significant CMB deviations~\cite{ru}.
Then, this constraint on $a_\xin$ can be propagated to get a
constraint on the value of $\varphi_*$ today, or equivalently on
$\alpha_{V0}$.

The result of this analysis is depicted in Figure~\ref{figVD}.
This figure can be understood as follows. First we see that almost
everywhere on the plane the solar system bound,
$\alpha_{V0}^2<10^{-5}$, is satisfied (the area above the
red line barely visible in the lower left corner); see Eq.~(\ref{alphalimit}).
Second, for a given value of $(\beta_V,\beta_D)$ we can read off
the maximum value of deviation from general relativity that can be
achieved in the Solar system once BBN constraints are satisfied.
The result from Ref.~\cite{couv} corresponds to the cut
$\beta_V=\beta_D$ of Fig.~\ref{figVD}. Figure~\ref{figVD} shows
clear structure with two sets of minima: (1) a set of parallel
lines with increasing periodicity and (2) a vertical line close to
$\beta_V\sim5$. Let us now try to understand this behavior.

First, the periodicity seen in this figure is not directly related
to the period~(\ref{pulssol}) of the time evolution of $\varphi_*$
but rather on the phase of this solution when evaluated at $p=0$.
Since we do not have an analytical solution in radiation-matter
era, as in Ref.~\cite{couv}, our discussion can only be
approximate but will still shed some light on our result. During the
radiation era, but after BBN, the field is frozen at a value
$\varphi_{\xout*}$. Assuming that at the time of matter-radiation
equality, $p_\xeq$, $\varphi_{\xeq*}\simeq\varphi_{\xout*}$, we
can obtain a solution for $\varphi_*$ at late times by matching
the solution~(\ref{ephisol}) to the constant solution at $p_\xeq$
so that
\begin{equation}
 \varphi_* = \varphi_{\xout*}\hbox{e}^{-\frac34(p-p_\xeq)}
 \cos\left[\frac34 r(p-p_\xeq) \right].
\end{equation}
It follows, then, that the value of the scalar field today is given by
\begin{equation}
 \varphi_{0*} = \varphi_{\xout*}\hbox{e}^{\frac34p_\xeq}
 \cos\left[\frac34 rp_\xeq \right],
\end{equation}
which illustrates that the oscillations of $\varphi_{0*}$, and
thus of $\alpha_{V_0}$, as a function of the two parameters
$(\beta_V,\beta_D)$, are due to the oscillations of $\varphi_{*0}$
as a function of $r$ and are thus determined by the phase at the
time of matching. To determine the points at which
$\varphi_{*0}=0$, we need the value of $p_\xeq$ that is determined
by imposing that $\rho_\mat + \rho_D = \rho_\rad$. This leads to
\begin{equation}
 \hbox{e}^{-p_\xeq} = \Xi_0\frac{\bar
 A(\varphi_{\xeq*})}{A_V(\varphi_{0*})}.
\end{equation}
The almost horizontal periodic structure in Fig.~\ref{figVD} is
related to the solution such that $\varphi_{0*}=0$. When this
happens, $\tilde\Xi_0=\Xi_0$ and $A_V(\varphi_{0*})=0$ and these
zeros occur when $\frac34 rp_\xeq= n \pi$ with $n\in\Bbb{N}$, that
is when
$$
\sqrt{\frac83\frac{\beta_V+\beta_D\Xi_0}{1+\Xi_0}-1} =
 \frac43 n\frac{\pi}{\ln\frac{\Omega_{\mat0}}{\Omega_{\rad0}}
 +\frac12 a_{V\xout} +\ln\left(1+\Xi_0\hbox{e}^{\frac{\beta_D-\beta_V}{\beta_V}a_{V\xout}} \right)}.
$$
Now, since the dark sector does not influence the dynamics during
electron-positron annihilation, $\varphi_{\xout*}$ depends
only on the values of $\varphi_{\xin*}$ and $\beta_V$. This
complicated relation is depicted in Fig.~2 of Ref.~\cite{couv}.
From this analysis, we argued that $a_{V\xout}$ is at most of
order $10^{-1}$-$10^{-2}$. This implies that we expect the
denominator of the r.h.s. of the previous expression to be
dominated by $\ln\frac{\Omega_{\mat0}}{\Omega_{\rad0}}$. Thus, the
positions in the plane ${\beta_V,\beta_D}$ for which
$\alpha_{V0}=0$ (and as a consequence $\alpha_{D0}=0$) corresponds
to a set of parallel lines $\mathcal{C}_n$ determined by
\begin{equation}
  \beta_V+\beta_D\Xi_0 = \frac38(1+\Xi_0)\left[\left(\frac{4 n\pi}{3
 \ln\frac{\Omega_{\mat0}}{\Omega_{\rad0}}}\right)^2 + 1 \right].
\end{equation}
This explains why the spacing between $\mathcal{C}_n$ and
$\mathcal{C}_{n+1}$ grows almost linearly and is mainly determined
by $\Xi_0$ and $\Omega_{\mat0}/{\Omega_{\rad0}}$, which arises
from the fact that the oscillations are related by the time of
matching. The slope is determined by $\Xi_0$ since it
characterizes the evolution during the matter era.

\begin{figure}[htb]
 \center\includegraphics[width=9cm]{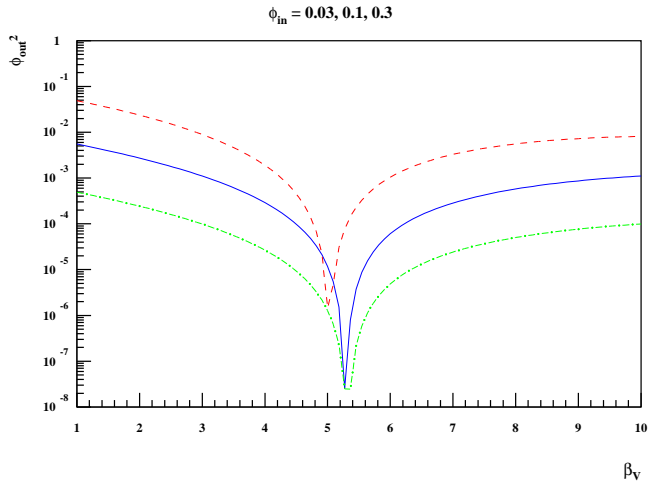}
 \caption{Evolution of $\varphi_{\xout*}$ as a function
 of $\beta_V$ assuming that $\phi_{\xin}=$ 0.03, 0.1 and 0.3.
 We see that the
 first minimum is obtained for $\beta_V\sim 5$, which corresponds
 to the vertical minimum of Fig.~\ref{figVD}. There is a slight dependence
 of the minimum with respect to $\phi_{\xin}$ and hence with the constraints
 provided by BBN.}
 \label{fig:3a-3b}
\end{figure}

Let us now turn to the second feature in Fig.~\ref{figVD}, that
is, the vertical line on which $\alpha_{V0}=0$. The relation
between $\varphi_\xout$ and $\varphi_\xin$ is not monotonous and
is determined mainly by $\beta_V$ (see Figure~2 of
Ref.~\cite{couv} that shows that it is periodic in $\beta_V$ with
a first minimum for $\beta_V\sim5$). Figure~\ref{fig:3a-3b} shows
how that the relation between $\varphi_\xout$ and $\varphi_\xin$
has indeed a minimum at $\beta_V\sim5$. This implies that for
these models $\varphi_*$ freezes to a very small value after the
electron-positron annihilation so that these models are more
efficiently attracted toward general relativity.

In summary, we can obtain a good understanding of our numerical
results and of the origin of the structures of the minima in
Fig.~\ref{figVD}. In particular, we found that the two sets of
structures do not have the same physical origin, one is related to
electron-positron annihilation and the second to oscillations in
the matter dominated era.

%%%%%%%%%%%%%%%%%%%%%%%%%%%%%%%%%%%%%%%%%%%%%%%%%%%%%%%%%%%%%%%%%%%%%
%%%%%%%%%%%%%%%%%%%%%%%%%%%%%%%%%%%%%%%%%%%%%%%%%%%%%%%%%%%%%%%%%%%%%

\subsubsection{Time variation of the gravitational constant}

From Eq.~(\ref{gdef}), using Eq.~(\ref{toto}), the time variation
of the gravitational constant is given by
\begin{equation}
 \sigma = 2\alpha_V\left(1+\frac{\beta_V}{1+\alpha_{V}^2}\right)
 \frac{\varphi_*'}{1+\alpha_V \varphi_*'},
\end{equation}
where $\sigma\equiv H^{-1}\dd\ln G_\cav/\dd t$ is constrained by
Eq.~(\ref{gconst0}), i.e. we must have $|\sigma_0|<\Sigma_0$,
where $\Sigma_0$ is the experimental upper bound on $\sigma$
today; see Eq.~(\ref{gconst0}).

\begin{figure}[htb]
 \center\includegraphics[width=8cm]{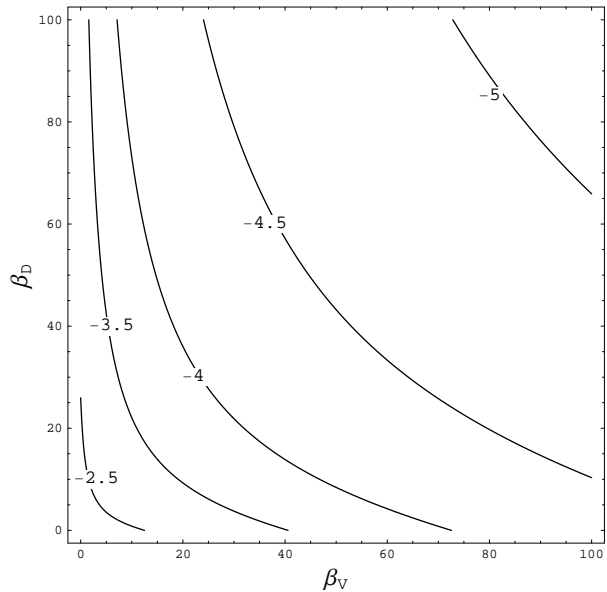}
 \caption{Contour plot of the time variation of the gravitational
 constant in the plane $(\beta_V,\beta_D)$ for different values of
 $\log\alpha_{V0}$ indicated on each curve. The model has to lie on the lower-left part
 of the plot so that $|\sigma_0|<\Sigma_0$.}
 \label{figGvar}
\end{figure}

Assuming we are in slow-roll today, we deduce that
$$
 (1+\Omega_{\Lambda0})\varphi'_{*0}\sim -\frac{\beta_V +
 \beta_D\Xi_0}{1+X_0}\varphi_0.
$$
We also deduce that the time variation of the gravitational constant
today is
\begin{equation}\label{s0}
 \sigma_0[\beta_V,\beta_D,\Xi_0,\alpha_{V0}]=2\alpha_{V0}\frac{\left(1+\frac{\beta_V}{1+\alpha_{V0}^2}\right)(\beta_V +
 \beta_D\Xi_0)}{(1+\Xi_0)(1+\Omega_{\Lambda0})-\alpha_{V0}^2(\beta_V +
 \beta_D\Xi_0)}.
\end{equation}
Figure~\ref{figGvar} shows the region of the parameter space
$(\beta_V,\beta_V)$ for which $|\sigma_0|<\Sigma_0$ given the
deviation from general relativity today, i.e. $\alpha_{V0}$. The
contour plots can be understood as follows. Since
$\alpha_{V0}^2\lesssim10^{-5}$, the
second term in the denominator of Eq.~(\ref{s0}) is negligible.
Unless we have $\beta$s of order $10^{5}$, we
conclude that
$$
 \frac{\sigma_0}{\Sigma_0 h}\sim 1.6\times10^{-2}\alpha_{V0}
 (1+ \beta_V)(\beta_V +
 \beta_D\Xi_0).
$$
This shows that even if the dynamics of the field is influenced by
$\beta_D$ which can be very large, the Solar system constraints on
$\alpha_{V0}$ are so strong that it requires very high $\beta_D$
to obtain a sizable time variation of the gravitational constant.

%%%%%%%%%%%%%%%%%%%%%%%%%%%%%%%%%%%%%%%%%%%%%%%%%%%%%%%%%%%%%%%%%%%%%
%%%%%%%%%%%%%%%%%%%%%%%%%%%%%%%%%%%%%%%%%%%%%%%%%%%%%%%%%%%%%%%%%%%%%

%%%%%%%%%%%%%%%%%%%%%%%%%%%%%%%%%%%%%%%%%%%%%%%%%%%%%%%%%%%%%%%%%%%%%%%%%%%%%%
\section{Conclusions}\label{sec5}

For a theory of gravity universally coupled to matter,
a scalar-tensor theory can be defined by a single coupling function $A(\varphi)$
and a scalar potential.  Such a theory can affect the evolutionary history of the
universe, particularly in models of extended quintessence.
Here we examined two types of quintessence models and examined the
effects of the non-minimal coupling on the evolution of the quintessence field.
For inverse power law quintessence models, we showed that the
evolution towards a tracking solution is accelerated due to mass
thresholds in the radiation dominated epoch, and during the subsequent matter
dominated epoch. In contrast, we saw that for quintessence models with defined minimum
which is also a minimum of the coupling function, the non-minimal coupling to gravity has
very little effect except at late times.

It is, however, possible that the dark sector of the theory
couples to gravity differently from that of the visible sector. In
this case, there is an additional coupling function as defined in
Eq.~(\ref{actionEF}). In this article, we have worked out the
cosmological evolution of a scalar-tensor theory with distinct
coupling functions for the coupling of the visible and dark
sectors of the theory to gravity. We developed the qualitative
features of the evolution with respect to the parameter space
defined by $\beta_V$ and $\beta_D$ which are derived directly from
the two coupling functions. We delineated those regions of the
parameter space which are attracted towards general relativity,
towards Brans-Dicke gravity, or towards a runaway solution. For
those theories with a late time attraction to general relativity,
we also derived additional constraints on the parameter space from
BBN and precision gravitational tests. In particular, it was shown
that in this class of models constraints more stringent than
those derived from the Solar system can be obtained from BBN. This
provides some insight on the equivalence principle between the
visible and dark sectors.

\section*{Acknowledgements}
The time consuming part of the computations was performed on the
computational grid GRIF (Grille de production pour la Recherche en
Ile de France, {\tt http://grif.fr}) and we thank Christophe
Diarra for his help during the implementation of the code. We
thank Carlo Schimd for his help in producing Fig.~\ref{figVD} and
Gilles Esposito-Far\`ese and Cyril Pitrou for discussions. The
work of KAO was partially supported by DOE grant
DE-FG02-94ER-40823.

%%%%%%%%%%%%%%%%%%%%%%%%%%%%%%%%%%%%%%%%%%%%%%%%%%%%%%%%%%%%%%%%%%%%%%%%%%%%%%
\appendix

\section{Numerical implementation}\label{appB}

From a numerical point of view, it is easier to integrate the
Einstein equations in the Einstein frame while using the cosmic
time $t$ in the  MJF. The expressions for the energy densities of the
visible sector are trivial in the MJF, where they can be
integrated analytically. It follows that the system reduces to
\begin{widetext}
\begin{eqnarray}
 && \frac{\dd\varphi_*}{\dd t} = A_V^{-1} \psi_*\label{s1} \\
 && \frac{\dd\psi_*}{\dd t} = -A_V^{-1}\left[3H_*\psi_*+ \frac{\dd V}{\dd\varphi_*}
 +4\pi G_*A_V^4(\alpha_D\rho_D + \alpha_\mat\rho_\mat)\right]\label{s2}\\
 && H^2_* = \frac{8\pi G_*}{3}A_V^4(\rho_D+\rho_\mat+\rho_\rad) + \frac{1}{3}\psi_*^2 +
 \frac{2}{3}V -\frac{K}{R^2_*}\label{s3}\\
 &&\frac{\dd\rho_\rad}{\dd t} + 4H\rho_\rad = 0,\label{s4}\\
 &&\frac{\dd\rho_\mat}{\dd t} + 3H\rho_\mat = 0,\label{s4b}\\
 && \frac{\dd\rho_D}{\dd t} + 3H\rho_D = A_{V}^{-1}\left(\alpha_D - \alpha_V \right)
 \rho_D\psi_*,\label{s5}\\
 &&H = A_{\rm V}^{-1}\left[H_* + \alpha_V\psi_*\right].\label{s6}
\end{eqnarray}
\end{widetext}
Note the asymmetry in the coupling functions, $A$, arises from the
fact that we have inserted the MJF energy densities rather than those defined in the EF.
The last equation is necessary to derive the Hubble
parameter in the MJF where the nuclear reactions are integrated.

\section{Local constraints}\label{appA}

The deviation from general relativity are constrained in the Solar
system. These deviations are usually summarized by constraints on
the post-Newtonian (PPN) formalism~\cite{will}. It is a general
formalism that introduces 10 phenomenological parameters to
describe any possible deviation from general relativity at the
first post-Newtonian order~\cite{will}. The formalism assumes that
gravity is described by a metric and that it does not involve any
characteristic scale. In our particular case, it is necessary that the
scalar field is light  so that a
Yukawa interaction on the scale of the solar system is not induced.

Since all of the matter in our Solar system stems from the visible
sector (i.e. the dark matter component of the Sun and planets is
supposed to be negligible), the two Eddington parameters (see
Refs.~\cite{will,gefp}) can be expressed in terms of the values of
$\alpha_V$ and $\beta_V$ today as
\begin{equation}
 \gamma^\ppn - 1 = -\frac{2\alpha_{V0}^2}{1+\alpha^2_{V0}},\qquad
 \beta^\ppn - 1 =\frac{1}{2}
 \frac{\beta_{V0}\alpha_0^2}{(1+\alpha_{V0}^2)^2}.
\end{equation}

Solar System experiments set strong limits on these parameters.
The perihelion shift of Mercury implies~\cite{mercurybound}
$|2\gamma^\ppn - \beta^\ppn -1|<3\times10^{-3},$ the Lunar Laser
Ranging experiment~\cite{llrbound} sets $4\gamma^\ppn - \beta^\ppn
-3 = -(0.7\pm1)\times10^{-3}$. Two experiments give a bound on
$\gamma^\ppn$ alone, the Very Long Baseline
Interferometer~\cite{vlbibound} sets $|\gamma^\ppn -1|<
4\times10^{-4}$ and the measurement of the time delay variation to
the Cassini spacecraft near Solar conjunction~\cite{cassinibound}
$\gamma^\ppn -1 = (2.1\pm2.3)\times10^{-5}$.

These two last bounds imply that $\alpha_{V0}$ has to be very
small, typically
\begin{equation}
 \alpha_{V0}^2<10^{-5}
 \label{alphalimit}
\end{equation}
while $\beta_{V0}$ can still be large~\cite{pulsar}. Binary pulsar
observations imply $\beta_{V0}\gtrsim-4.5$. Note that even
though $\beta_0$ is not bounded above by experiment, we will
assume that it is not very large, typically we assume
$\beta_{V0}\lesssim100$, so that the post-Newtonian approximation
scheme makes sense.
Note that none of these observations constrain the dark sector and
$(\alpha_{D0},\beta_{D0})$ are completely free.

Since Cavendish experiment and planets are composed only of
baryonic matter, we deduce that the gravitational constant is
given by
\begin{equation}\label{gdef}
 G_\cav = G_*A_V^2(1+\alpha_V^2).
\end{equation}
Using $p$ as the time variable as defined in Eq.~(\ref{def_p}), the time
variation of the gravitational constant today determines the time
derivative of $\varphi_*$ today
\begin{equation}\label{gconst}
 2\alpha_{V0}\left[1+\frac{\beta_{V0}}{1+\alpha_{V0}^2}-\frac{\Sigma_0}{2} \right]
 \left.\frac{\dd\varphi_*}{\dd p}\right|_0=\Sigma_0\ .
\end{equation}
Experimentally, this is bounded~\cite{dickey} by
\begin{equation}
 \frac{1}{G_\cav}\frac{\dd G_\cav}{\dd t} = \Sigma_0 H_0,\qquad
 |\Sigma_0| < 5.86\times10^{-2}h^{-1}.
 \label{gconst0}
\end{equation}
The new main feature compared to the models analyzed in
Ref.~\cite{couv} is that the dynamics of the scalar field depends
on the dark sector coupling and thus on $\alpha_D$.

\section{BBN constraints}\label{appC}

BBN is one of the most sensitive available probes of the very
early Universe and of physics beyond the standard model. Its
success rests on the concordance between the observational
determinations of the light element abundances of D, \he3, \he4,
and \li7, and their theoretically predicted abundances \cite{cfo1,cvcr01}.
Furthermore, measurements of the CMB anisotropies by WMAP
\cite{obs2,dunkley} have led to precision determinations of the baryon
density or equivalently the baryon-to-photon ratio, $\eta$.
The new WMAP 5-year data alone is
$\Omega_\baryon h^2$ = 0.02273$\pm$0.00062
 and is equivalent to
$\eta_{\rm 10,{\rm CMB}}$ = 6.23$\pm$0.17, where $\eta_{10}$ =
$10^{10} \eta$.
Using the WMAP data to fix the baryon density,
the light element abundances \cite{cfo3,coc,cyburt,coc2,cuoco} can be
quite accurately predicted.
The wealth of the cosmological data are obtained from
spectroscopic observations and compared
directly with BBN predictions assuming the WMAP determination of
$\Omega_\baryon h^2$.

Note that the \he4 abundance is often used as a sensitive
probe of new physics ( see e.g. \cite{cfos}).  This is due to the fact that nearly all
available neutrons at the time of BBN end up in \he4 and the
neutron-to-proton ratio is very sensitive to the competition between
the weak interaction rate and the expansion rate.

In the following, we briefly state the abundance measurements used in our analysis.

\subsubsection{D/H}

The best determinations of primordial D/H are based on
high-resolution spectra in high-redshift, low-metallicity quasar absorption systems (QAS),
via its isotope-shifted Lyman-$\alpha$ absorption.
The seven most precise observations of deuterium
in QAS give (see \cite{Pettini} and references therein) a weighted
mean value of D/H = $(2.82 \pm 0.21) \times 10^{-5}$
(1--$\sigma$) in good agreement with BBN at the WMAP value of $\eta$.
Note that the uncertainty quoted above is purely statistical and there remains
considerable scatter in the data.

\subsubsection{\he4}

\he4 is observed  in clouds of ionized hydrogen (HII regions), the
most metal-poor of which are in dwarf galaxies. There is a
large body of data on \he4 in these systems~\cite{iz,iz2}
for which an extended data set including 89 HII regions obtained
$Y_p$ $=$ 0.2429 $\pm$ 0.0009~\cite{iz2}.  However, the
recommended value is based on the much smaller subset of 7 HII
regions, finding $Y_p$ $=$ 0.2421 $\pm$ 0.0021.

\he4 abundance determinations depend on several
physical parameters associated with the HII region in addition to
the overall intensity of the He emission line.  These include, the
temperature, electron density, optical depth and degree of
underlying absorption. Unfortunately, there are severe degeneracies inherent in the
in the determination of the \he4 abundance \cite{OSk}.
Using a subset of the highest quality data from the sample of Izotov and
Thuan~\cite{iz}, Monte Carlo methods were used to extrapolate the  \he4
abundance \cite{os2} which was determined to be $Y_p = 0.2495 \pm 0.0092$.
Conservatively, it would be difficult at this time to exclude any
value of $Y_p$ inside the range 0.232 -- 0.258.

\subsubsection{\li7/H}

The systems best suited for Li observations are metal-poor halo
stars in our Galaxy.  Analyses of the abundances in these stars
yields \cite{rbofn} ${\rm Li/H}|_p = (1.23^{+0.34}_{-0.16}) \times
10^{-10}$ when systematic uncertainties are included.
The \li7 abundance based on the WMAP baryon density is predicted
to be~\cite{coc}
\li7/H $= 4.15^{+0.49}_{-0.45} \times 10^{-10}$.
We note that a recent reanalysis of the
$\he3(\alpha,\gamma)\be7$
reaction, which is the most important \li7 production process in BBN,
was considered in detail in \cite{cd}. When the new rate is used a
higher \li7 abundance is found \cite{cfo5}
 \li7/H $= (5.24^{+0.71}_{-0.62}) \times 10^{-10}$.

%%%%%%%%%%%%%%%%%%%%%%%%%%%%%%%%%%%%%%%%%%%%%%%%%%%%%%%%%%%%%%%%%%%%%%%%%%%%%%

\end{document}